\def\ap{\mbox{\bf{astro-ph}}}
\def\aap{\mbox{\bf{Astron.\ Astrophys.}}}
\def\apj{\mbox{\bf{Astrophys.\ J.}}}
\def\apjl{\mbox{\bf{Astrophys.\ J.\ L.}}}
\def\aj{\mbox{\bf{Astron.\ J.}}}
\def\aaps{\mbox{\bf{Astron.\ Astrophys.\ Suppl.}}}

\def\pasp{\mbox{\bf{Pub.\ Astron.\ Soc.\ Pacific}}}
\def\mnras{\mbox{\bf{Mon.\ Not.\ Roy.\ Astron.\ Soc.}}}
\def\ap{\mbox{\bf{astro-ph}}}

\def\araa{\mbox{\bf{Ann.\ Rev.\ Astron.\ Astrophys}}}
\def\apjs{\mbox{\bf{Astrophys.\ J.\ Suppl.}}}

\newcommand{\msun}{\mbox{$\mbox{M}_{\odot}$}}

\newcommand{\reff}{\mbox{$r_{\rm eff}$}}

\newcommand{\kms}{\mbox{km$\,$s$^{-1}$}}
\newcommand{\nrextended}{\mbox{305}}
\newcommand{\besttfour}{\mbox{$1.0^{+0.6}_{-0.5}\times10^8$}}

\newcommand{\besttfourwithgamma}{\mbox{$1.0^{+0.84}_{-0.35}\times10^8$}}
\newcommand{\besttfourwithinc}{\mbox{$2.0^{+5.2}_{-1.1}\times10^8$}}
\newcommand{\besttfourwithstep}{\mbox{$2.0^{+2.3}_{-1.1}\times10^8$}}

\newcommand{\bestgamma}{\mbox{$0.65_{-0.25}^{+0.16}$}}
\newcommand{\bestcfrinc}{\mbox{$7.0^{+68.1}_{-5.0}$}}
\newcommand{\bestcfrstep}{\mbox{$3.0^{+4.6}_{-1.2}$}}

\newcommand{\redchi}{\mbox{$\chi_\nu^2$}}

\documentclass{aa}
\usepackage{graphicx}

\begin{document}
   \title{The Star Cluster Population of M51: III. Cluster disruption
   and formation history}


   \author{M. Gieles \inst{1} \and N. Bastian \inst{1} \and
          H.J.G.L.M. Lamers \inst{1,2} \and J.N. Mout \inst{1} 
          }

   \offprints{gieles@astro.uu.nl}

   \institute{$^1$Astronomical Institute, Utrecht University, 
              Princetonplein 5, NL-3584 CC Utrecht, The Netherlands \\
              \email{gieles@astro.uu.nl} \\
             $^2$SRON Laboratory for Space Research, Sorbonnelaan 2, 
             NL-3584 CA Utrecht, The Netherlands\\
             }

   \date{Received \today; accepted ??, 2005}


   \abstract{In this work we concentrate on the evolution of the
cluster population of the interacting galaxy M51 (NGC~5194), namely the timescale
of cluster disruption and possible variations in the cluster formation
rate. We present a
method to compare observed age vs. mass number density diagrams with
predicted populations including various physical input parameters like
the cluster initial mass function, cluster disruption, cluster
formation rate and star bursts. If we assume that the cluster formation rate
increases at the moments of the encounters with NGC~5195, we find an
increase in the cluster formation rate of a factor of $\bestcfrstep$,
combined with a disruption timescale which is slightly higher then when assuming a constant formation rate ($t_4 =
\besttfourwithstep$ yr vs. $\besttfour$ yr). The measured cluster
disruption time is a factor of 5 shorter than expected on
theoretical grounds. This implies that the disk of M51 is not a
preferred location for survival of young globular clusters, since even
clusters with masses of the order of $10^6 \msun$ will be destroyed
within a few Gyr.  

}
         
\maketitle

\section{Introduction}
\label{sec:introduction}

The goal of this series of papers is to understand the properties of
the entire star cluster population of the interacting spiral galaxy
M51.  These properties include the age and mass distribution of the
cluster population.  Additional properties are the survival rate of
the clusters as well as any relations between the observed
properties. These relations may be used to constrain cluster formation
and destruction scenarios.

In order to study the above properties, we exploit the large amount of
{\it HST} broad-band archival data on M51, which covers roughly 50\%
of the observed surface area of M51, and covers a broad spectral range
({\it UV} to {\it NIR}).  The large spatial coverage is necessary in
order to obtain a large sample of clusters for carrying out a
statistical analysis, and the broad spectral range allows accurate
determinations of the individual cluster properties (Bik et al. 2003;
Anders et al. \cite{anders04}).  A preliminary analysis of a subset of
the M51 cluster population was carried out by Bik et al. (2003;
hereafter Paper I) who introduced the method used to determine the
cluster properties and derived the age and mass distributions of the
cluster sample roughly 2 kpc to the North East of the nucleus.

Bastian et al. (\cite{bastian05}; hereafter Paper II) extended the survey to
include the entire inner $\sim5$ kpc of M51, and found 1152 clusters,
$\nrextended$ of which had accurate size determinations.  In that work we
extended the age distribution analysis of Paper I and found evidence
for a cluster formation rate increase $\sim 50-70$ Myr ago. This
corresponds to the last close passage of NGC~5195 and M51 (Salo \&
Laurikainen \cite{salo00}).  Additionally we found that 68 $\pm 15$\%
of the clusters forming in M51 will disrupt within the first $\sim 10$
Myr after their formation, independent of their mass, so-called {\it
infant mortality}.  For the resolved cluster sample, we found that the
size distribution (the number of clusters as a function of their
effective radius) can be well fit by a power-law: $N\,d\reff
\propto \reff^{-\eta}\,d\reff$, with $\eta = 2.2 \pm 0.2$,
which is very similar to that found for Galactic globular clusters.
Finally, we did not find any relation between the age and mass, mass
and size, nor distance from the galactic center and cluster size.

In this study we focus on the evolution of the population of clusters
in M51, in particular the timescale of cluster disruption and possible
variations in the cluster formation rate. Cluster disruption of
multi-aged populations, which excludes the galactic globular clusters,
has been the subject of many earlier studies (e.g. Hodge
(\cite{hodge87}) for the SMC and Battinelli \& Capuzzo-Dolcetta
(\cite{battinelli91}) for the Milky Way). In this study we will take
into account mass dependent disruption, since the time needed to
destroy half of the cluster population, which has been estimated in
earlier work, will strongly depend on the mean mass of the sample and
the lower mass limit of the sample. In addition, we here want to study
the effect of variations in the formation rate, which is usually kept
constant.

Boutloukos \& Lamers (\cite{boutloukos03}) have
developed a method to derive the disruption timescale based on the age
and mass distributions of a magnitude limited cluster sample. They
found that the disruption time of clusters in M51 is a factor of 15
shorter than the one for open clusters in the solar
neighborhood. Lamers, Gieles
\& Portegies Zwart (\cite{lamers05a}) showed that part of the
difference can be explained by the difference in density of the
cluster environment. They showed that the disruption time of clusters
depends on the clusters initial mass and the galaxy density as $t_{\rm
dis} \propto M_i^{0.62}\,\rho_{\rm gal}^{-0.5}$, based on the results
of $N$-body simulations. The disruption time of clusters in M51 was
still about a factor of 10 lower than the predicted value. In this
work we are particularly interested if a short disruption timescale
can be mimicked by an increasing cluster formation rate, and how the
assumed disruption law influences the derived timescales.  To this end
we have generated artificial cluster samples with parameterized global
characteristics (e.g. time dependent cluster formation rates,
disruption laws, infant mortality rates, and mass functions).  We then
compare these models with the derived age and mass distributions of
the cluster population of M51 to derive the best fit parameters for
the population as a whole.

The structure of the paper is as follows: In \S~\ref{sec:observations}
the observations of the cluster population of M51 are presented. In
\S~\ref{sec:parameters} we investigate how the disruption time depends
on different cluster and galaxy parameters. \S~\ref{sec:explanation}
describes the steps we will take in our models, where the details of
the models we used to generate artificial cluster populations will be
explained in \S~\ref{sec:models}.  The results of the fits are given in
\S~\ref{sec:fits}. A discussion on the implication of the results is given in
\S~\ref{sec:implications}. The conclusions are presented in
\S~\ref{sec:conclusions}.

\section{The observations}
\label{sec:observations}

\subsection{Fitting the observed spectral energy distribution}
\label{subsec:fitting}

From archival {\it HST} broadband photometry we have derived the age,
mass, and extinction of 1152 clusters in M51 (Paper II) using the
three-dimensional maximum likelihood fitting (3DEF) method. Details
about the 3DEF method can be found in Paper I. In summary, the
spectral energy distribution of each cluster is compared with cluster
evolution models. In this case the {\it GALEV} simple stellar
population (SSP) models (Anders et al. \cite{anders03}; Schulz et
al. \cite{schulz02}) for solar metallicity and Salpeter IMF are
used. For each age a series of different extinctions is then applied
to the models and all combinations of age and extinction are compared
to the data. The lowest $\chi^2$ is kept and from the absolute magnitude at
that age the mass is determined. Detailed tests of the accuracy and
reliability of the derived parameters are presented in Paper II.
In the present work we further investigate the accuracy of our fitting
method and we use the data set to develop a model that describes the
global properties of the cluster system.

\subsection{The age and mass distribution of clusters}
\label{subsec:agemass}

\begin{figure}[h]

    \includegraphics[width=8cm]{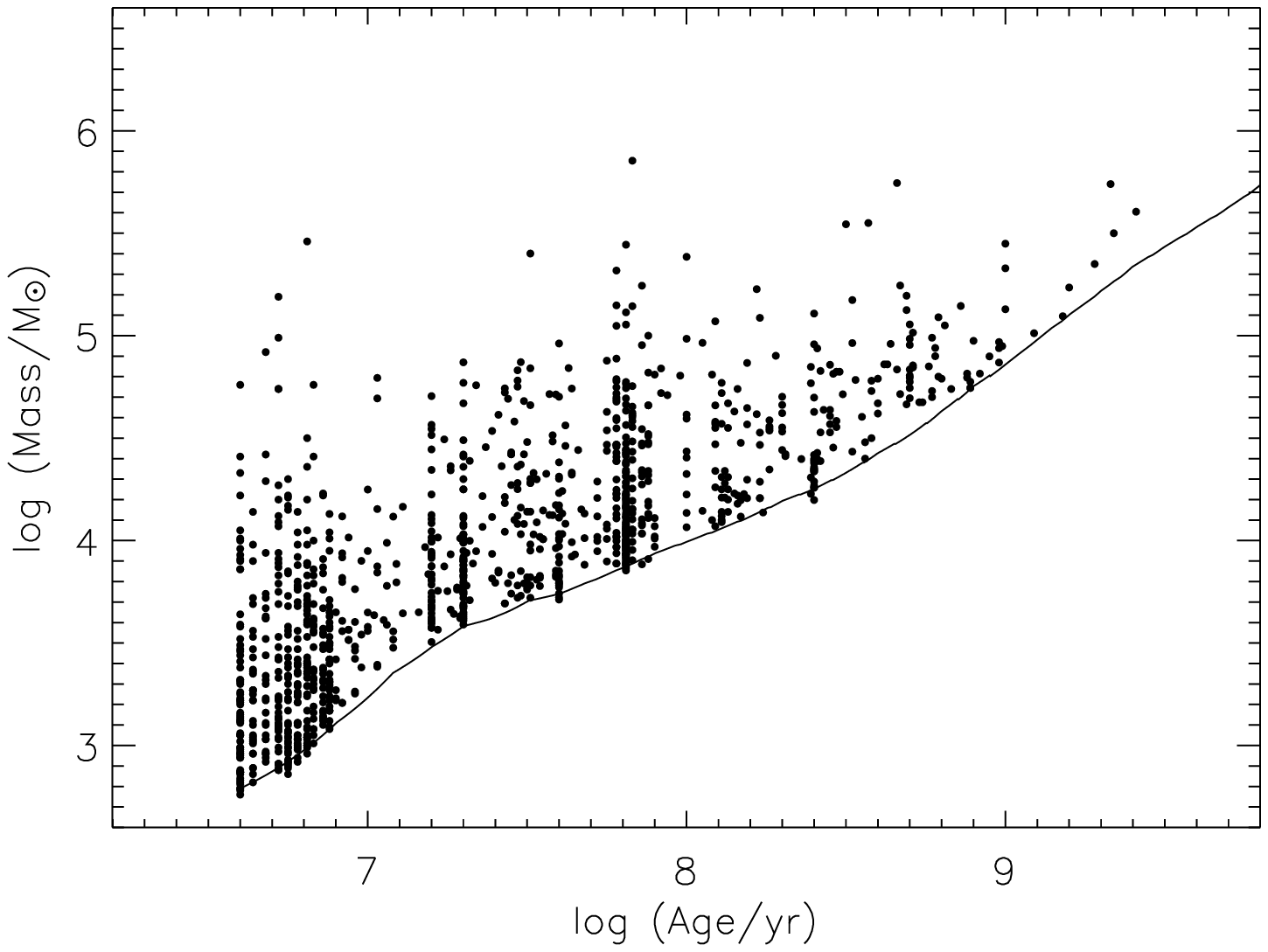}
    \includegraphics[width=8cm]{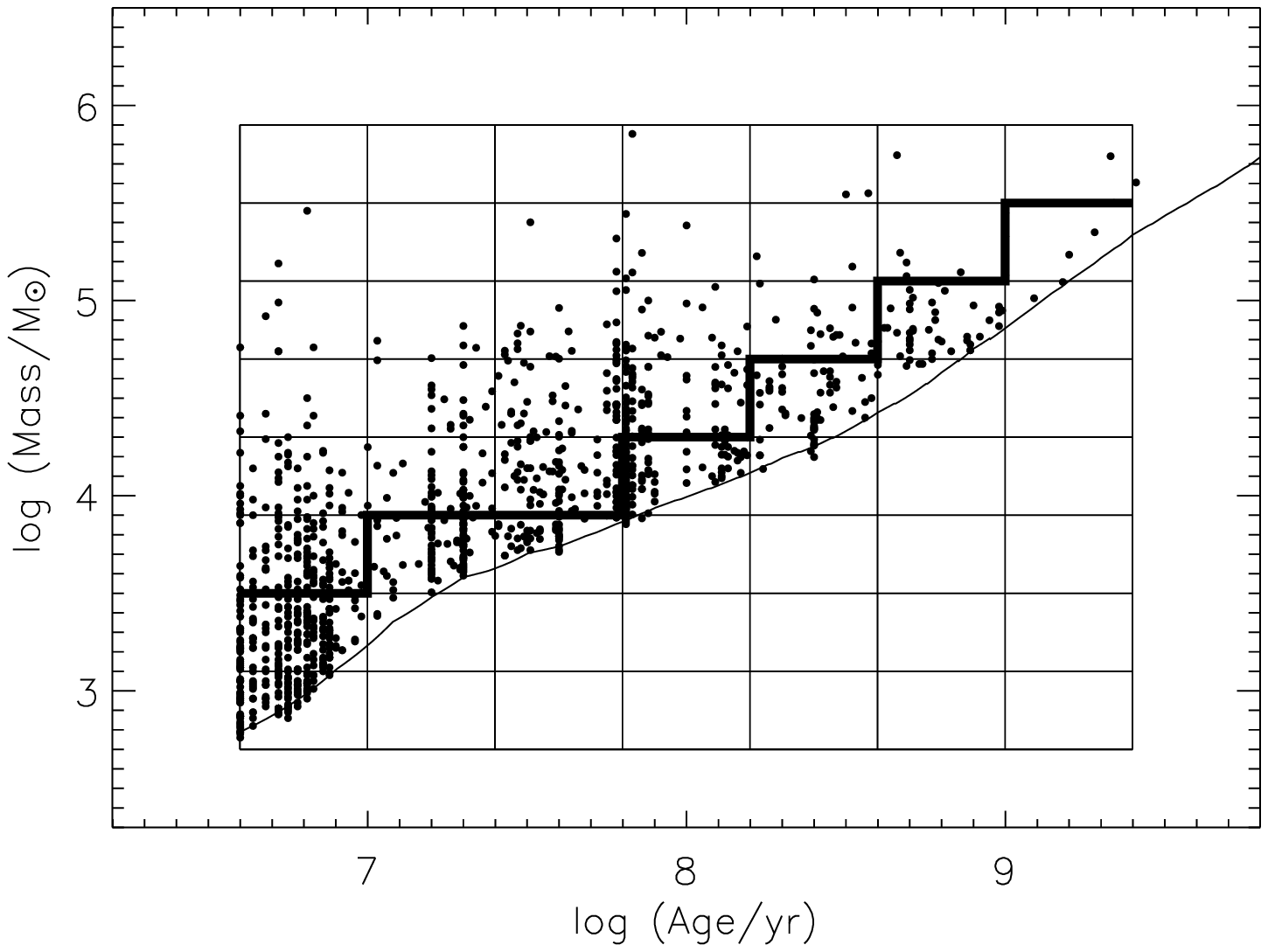}
    \includegraphics[width=8cm]{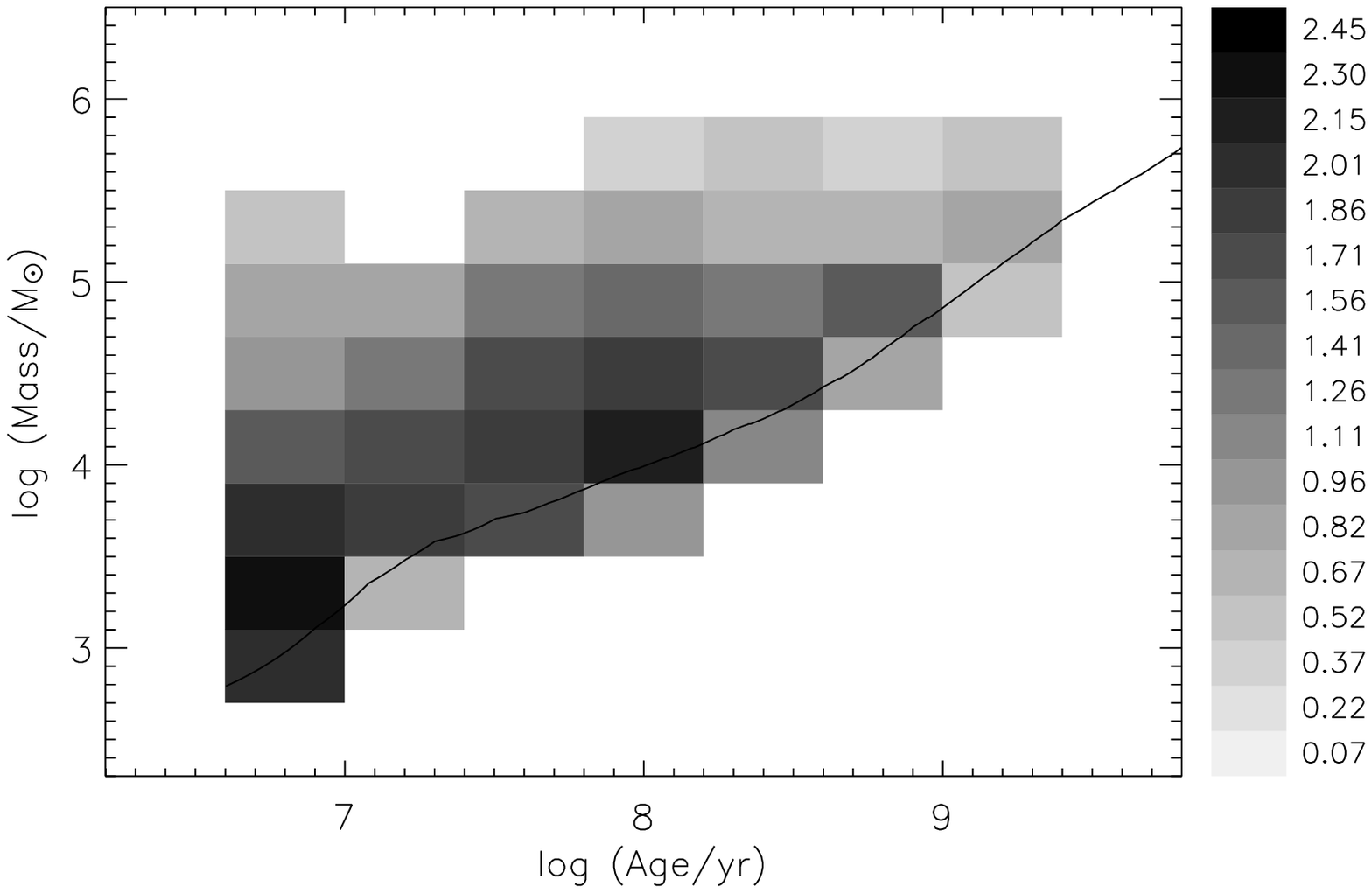}

    \caption{Ages and present masses of the 1152 clusters identified
    in Paper II. {\bf Top}: original data, where every point
    represents a cluster. {\bf Middle:} same data as in top panel,
    with overplotted the grid used to bin the data. Above the dark
    line are the bins which are not affected by the detection
    limit. {\bf Bottom}: logarithmic density plot of the same sample,
    where dark regions represent more clusters. The right hand scale
    shows how the grey values correspond to the logarithm of
    number. The line in all three plots is the 90\% completeness limit (F439W =
    22.6 mag).}

    \label{fig:age-mass}
                                                                                                                   
\end{figure}

The ages and present masses of the 1152 clusters are plotted in the
top panel of Fig.~\ref{fig:age-mass}. In order to be able to compare
our observations with simulated cluster samples, we bin the data in
logarithmic number density plots of the age vs. mass
distribution. Clusters where counted in bins of 0.4 age dex by 0.4
mass dex (Fig.~\ref{fig:age-mass}, Middle). The result is illustrated
in the bottom panel of Fig.~\ref{fig:age-mass}. A few striking features can be
learned from this diagram:

\begin{enumerate}
\item a burst in the cluster formation rate (CFR) between 50 and 70 Myr,
corresponding with the most recent interaction with the companion
galaxy NGC~5195;
\item a short lived young population with ages $<$ 10 Myr. In Paper II we
found that $\pm$68\% of these young clusters will dissolve within 10
Myr, independent of their mass;
\item evolutionary fading under the detection limit, which makes it harder
to detect old low mass clusters. The increasing line in
Fig.~\ref{fig:age-mass} shows how the 90\% completeness limit in the
F439W band (22.6 mag.) corresponds with different masses at different ages;
\item an {\it apparent} increase in the mass of the most massive cluster
with age. This is a binning effect: the older age bins span more time
and therefore contain more clusters and the chance of finding a more
massive cluster at older ages is higher due to the {\it size of
sample} effect (Hunter et al. \cite{hunter03}). We have tried to use
the method of Hunter et al. (\cite{hunter03}) to derive the cluster
formation rate, but we found that the increase in the maximum mass is
much to shallow. This probably means that M51 has reached the maximum
cluster mass ($\sim 10^6 \msun$) and then the relation of Hunter et
al. (\cite{hunter03}) does not apply anymore, since the maximum
cluster mass found at a certain age is then not determined anymore by
sampling statistics. This is the topic of a next study
(Gieles et al.~\cite{gieles05}).

\end{enumerate}

We found that a large fraction ($\pm 68$\%) of the clusters younger
than $10^7$ yr dissolves independent of mass, probably due to the
removal of the primordial gas (e.g. Kroupa \cite{kroupa04}; Geyer \&
Burkert \cite{geyer01}; Lada \& Lada \cite{lada03}). After these critical 10$^7$ years, the
surviving clusters will dissolve due to the tidal field of the host
galaxy and external perturbations from, for example, encounters with
giant molecular clouds (GMCs). The disruption of clusters is in that
sense a two step process. Here we will use the resulting age/mass
distribution to study the disruption of clusters with ages larger than
$10^7$ yr., i.e. the second step in the disruption process.

\subsection{Artifacts introduced by the age fitting method}
\label{sec:artifacts}

We want to see whether the 3DEF method (\S~\ref{subsec:fitting}), used
to derive ages, masses and extinctions from the photometry introduces
systematic artifacts. More important, could it affect our results of
the disruption time or formation rate?  For instance, are there systematically old clusters
fitted with young ages or the other way around?

The uncertainty in the derived ages, extinctions and masses from
broad-band photometry is mainly caused by two effects:

\begin{enumerate}
\item Differences between the real integrated colors of the clusters and the
models, for example due to stochastic sampling of the stellar IMF
which cannot be taken into account in the SSP models, or errors in the
stellar isochrones used;

\item Systematic errors introduced by the age fitting method and the applied
selection effects
\end{enumerate}

Of course the first effect cannot be corrected for, unless we are able
to compare photometric determined ages directly with spectroscopically
determined ages, which is unfortunately not feasible for a whole
population of clusters. In addition, although spectroscopically
derived ages are more accurate, they are hampered by their own
problems (Brodie et al. \cite{brodie98}). In addition, the age
derivation will dependent on the choice of the adopted SSP
models. Variations in the metallicity and the IMF will affect the
derived ages. A detailed comparison of the data with models of
different metallicity is carried out in Paper II. An earlier study of
the age distribution of M51 used Starburst99 models (Bastian \& Lamers
\cite{bastian03}), and they found a very similar age distribution.

The second effect can be quantified with the use of
artificial cluster populations. Earlier studies (e.g. Anders et
al. \cite{anders04}; de Grijs et al. \cite{degrijs05}) have already
shown the importance of using a long wavelength baseline ($U$ to
$NIR$) for age dating young clusters. Here we will make an attempt to
quantify possible systematic errors introduced by the age-fitting
method and see whether we can correct for them or not.

To quantify the artifacts introduced by the fitting routine, an
artificial cluster sample including simulated observational errors and
extinction values is generated and fitted with the same fitting
procedure as used for the data (\S~\ref{subsec:fitting}). We start
with a sample of clusters equally spread in log(Age/yr) and
log($M/\msun$) space. In total 201 time steps between log(Age/yr) = 6
and log(Age/yr) = 10 and 161 mass-steps between log($M/\msun$) = 2 and
log($M/\msun$) = 7 were generated. The {\it GALEV} models have
log(Age/yr) = 6.6 as youngest model, so clusters with younger ages
were given that age. The magnitudes as a function of age and mass
where taken from the {\it GALEV} SSP models. Observational
uncertainties were applied as a function of magnitude as was done in
Paper II: the observed errors in the magnitudes of clusters in M51 can
be well approximated by $\Delta {\rm mag}_\lambda = 10^{d_1 +
d_2\times {\rm mag}_\lambda}$. The values for $d_1$ and $d_2$ for the
different filters are results from analytical functions fitted to the
observed errors and magnitudes and are given in Table~4 of Paper
II. Ideally, we then apply the same extinction to the model clusters
as the M51 clusters have. Unfortunately the only information we have
is the extinction we have measured, which of course could already be
polluted with artifacts. To get an estimate of the uncertainty in the
measured extinction, we start with a sample of clusters with no
extinction applied. When we fit this population with the 3DEF method,
we find that 20\% of the sources is fitted with some extinction. This
is quite a large number, but fortunately 90\% of these sources have
extinction values lower than $E(B-V)$ = 0.1 mag. The maximum $E(B-V)$
found is 1 mag.

The next step is to apply an extinction model close to what we
observe. To this end $E(B-V)$ extinction values were chosen randomly
from a Gaussian distribution centered at 0 with $\sigma = 0.10$ for
clusters younger than log(Age/yr) = 7.3 and $\sigma = 0.05$ for
clusters older than log(Age/yr) = 7.3. The values for $\sigma$ agree
with the value we found for the mean extinction in Paper II. There we
found that these values are the average extinction for these two age
groups. The higher extinction for young ages is caused by the presence
of the left-over dust around the cluster. Negative extinctions were
set to 0, resembling the extinction distribution of the data where
half of the clusters had $E(B-V)$ = 0 (See Fig.~8 of Paper II). An age
dependent maximum extinction was applied of the form: $E(B-V)_{\rm
max}(t) = 5 -0.5*\log(t)$. This is a little bit lower than the
observed maximum extinction, but we know that some of the observed
high values could be caused by wrong fits. This still resembles the
observed extinction behavior quite well. The resulting magnitudes are
than cut off at our completeness limits in each filter. In this way we
created the spectral energy distributions of a large artificial
cluster sample with age, mass and extinction known for each
cluster. These were fitted with the 3DEF method
(\S~\ref{subsec:fitting}).

\begin{figure}
    \includegraphics[width=8cm]{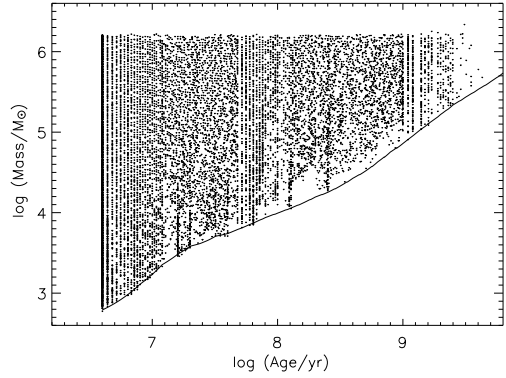}

 \caption{Age/mass diagram of artificial sample after fitting with the
     method described in \S~\ref{subsec:fitting}. The deviations from
     equally spaced dots in log(Age/yr)/log($M/\msun$) are caused by the
     applied observational errors and the fitting routine.}

    \label{fig:fitsim}
    \end{figure}

\begin{figure}
     \includegraphics[width=8.5cm]{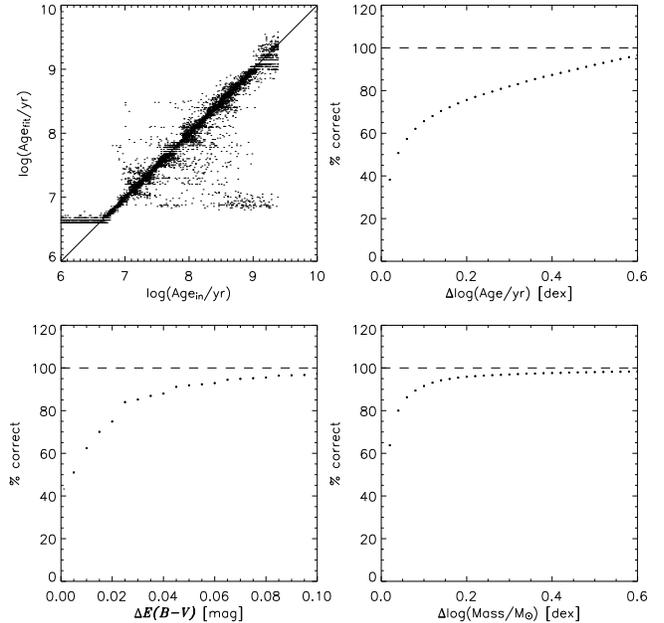}

      \caption{Result of fitting an artificial cluster sample with the
 3DEF method. {\bf Top left:} The fitted age is shown versus the input
 age for each cluster. Deviations from the one-to-one relation are
 caused by photometric errors which are applied to the input sample
 and misfitted extinction.{\bf Top right:} The percentage of clusters
 fitted with the same age as the input value plus some deviations, as
 a function of this deviation. {\bf Bottom left:} similar but then for
 extinction. {\bf Bottom right:} similar but then for the mass.}

    \label{fig:ageout-agein}
    \end{figure}

The result of the fitted simulation is shown in
Fig.~\ref{fig:fitsim}. A direct comparison with the observed age-mass
diagram of M51 clusters (Fig.~\ref{fig:age-mass}, Top) shows that
there are features present in the data which are not visible in the
fitted simulation. For example, there is a gap at 6.9 $<$
log(Age/yr) $<$ 7.1 in the M51 cluster sample which seems to appear at
slightly higher ages in the fitted simulations (7.1 $<$ log(Age/yr)
$<$ 7.2 ). This suggests that the artifacts in the data are not only
caused by our applied selection effects or our age-fitting
technique.  In the top left panel of Fig.~\ref{fig:ageout-agein} we
show the fitted age versus the input age for the simulated cluster
sample. A large number of clusters with wrong ages are fitted with an
log(Age/yr) $\simeq$ 7. We found for 87\% of the modeled clusters that
the fitted age was the same as the input age within 0.4 dex
(Fig.~\ref{fig:ageout-agein}, Top right). For the mass 97\% was fitted
correctly within 0.4 dex (Fig.~\ref{fig:ageout-agein}, Bottom right)
and 92\% of the extinction values where fitted back within 0.05
mag. (Fig.~\ref{fig:ageout-agein}, Bottom left). We have to realize
that the strength of the artifacts depend on the number of input
clusters at each age and mass bin. We have not attempted to match the
observations in this stage, since we are only interested in relative
errors. For example, the horizontal spur at log(Age/yr) $\simeq$ 7 in
Fig.~\ref{fig:ageout-agein} (Top, left) is populated with clusters
with input ages up till a few times 10$^9$ yr. The number of clusters
with that age in our M51 sample is very low (See
Fig.~\ref{fig:age-mass}, Top).

\subsection{Correcting the data for fitting artifacts}
\label{subsec:correct}
\begin{figure}[t]
    \includegraphics[width=8cm]{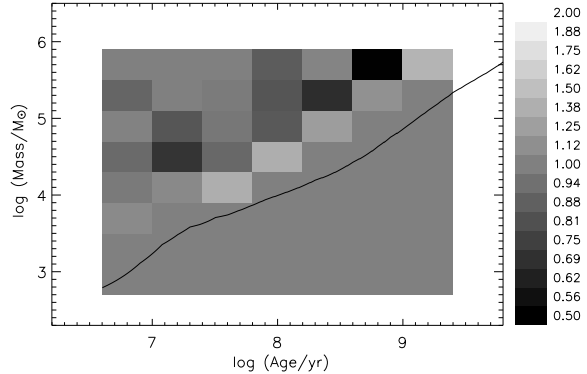} \caption{Ratio of the
    corrected observations over the uncorrected observations. The
    corrected data was calculated using Eq.~\ref{eq:contribution} and
    Eq.~\ref{eq:correct}. Light regions indicate where less
    clusters are found back by the fitting procedure. Dark regions
    indicate where more clusters are fitted than inputted.}

    \label{fig:ratio}
\end{figure}

We will try to correct the data for possible artifacts, by
using the (systematic) deviations found in
\S~\ref{sec:artifacts}.  Correcting the observed ages based on the
absolute numbers deviating from the one-to-one relation is not useful,
since the number of clusters that were used as input at each age and
mass differs from the observed number. From the input sample we can
derive how many clusters are (systematically) fitted with wrong ages
and masses. Let the total number of bins in age/mass space be
$K$. Here we define the number of bins as the number of bins which are
not affected by the detection limit (See Fig.~\ref{fig:age-mass},
Middle panel). The number of clusters found in each bin is the sum of
the contribution of clusters from all bins to this one, where the
majority will be from the bin with the same input age and mass. When
we write the number of clusters in each bin as a vector with
$K$-entries, the fitted number of clusters can be written as a matrix
multiplication of all contributions times the input number of clusters

\begin{equation}
\left( \begin{array}{c}
N_1 \\ N_2 \\ \vdots \\ N_K\\
\end{array} \right)_{\rm obs}
=
\left( \begin{array}{cccc}
C_{11} & C_{12} & \ldots & C_{1K} \\
C_{21} & C_{22} & \ldots & C_{2K} \\
\vdots & \vdots & \ddots & \vdots \\
C_{K1} & C_{K2} & \ldots & C_{KK} \\
\end{array} \right)
\left( \begin{array}{c}
N_1 \\ N_2 \\ \vdots \\ N_K\\
\end{array} \right)_{\rm intr}
\label{eq:contribution}
\end{equation}
where $N_{{\rm intr},j}$ is the number of clusters generated in bin $j$,
$N_{{\rm obs},i}$ is the number of clusters fitted in bin $i$ and $C_{ij}$ is
the contribution to bin $i$ from bin $j$. Tests have shown that the
best results are acquired when only taking bins into account which are
not affected by the detection limit (see Fig.~\ref{fig:age-mass},
Middle panel). From the simulated and fitted sample we can derive the
values for $C_{ij}$ for all combinations of $i$ and $j$. All values on
the diagonal of $C$ (i.e. $C_{ij}$ where $i$ = $j$) are close to 1
since most clusters are fitted with the same age, extinction and
mass. All other values are 0 or between 0 and 1. When we know the
matrix $C$, the inverse can be used to correct the observations for
systematic 3DEF fitting artifacts.

\begin{equation}
\vec{N}_{\rm intr} = C^{-1}\times\vec{N}_{\rm obs}
\label{eq:correct}
\end{equation}
where $\vec{N}_{\rm intr}$ is the vector with the intrinsic number of
clusters, $C^{-1}$ is the inverse of the contribution matrix as
defined in Eq.~\ref{eq:contribution} and $\vec{N}_{\rm obs}$ is the
vector with observed clusters. When we correct the observed vector and
plot it in a 2D age-mass diagram again, divide the corrected and
uncorrected observation, we can see where deviations take
place. Fig.~\ref{fig:ratio} shows the ratio of corrected over
uncorrected observations. The number of clusters in the age bins at
log(Age/yr) = 7.2 and 8.0 has been lowered with about 15\%. The
observations, however, show a gap at log(Age/yr) = 7.2
(Fig.~\ref{fig:age-mass}, Top). So, the underestimation of clusters in
that age bin is not caused by our fitting routine. 

\begin{figure}[t]
    \includegraphics[width=8cm]{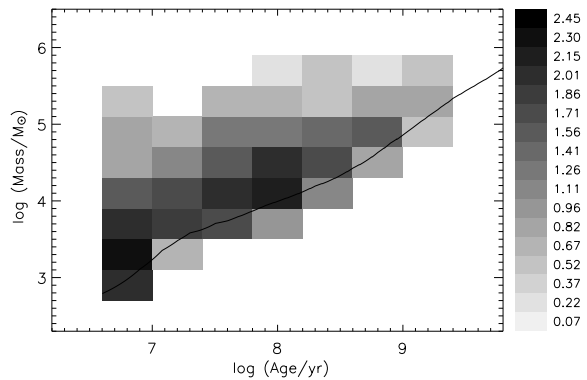} 

    \caption{Corrected age and mass distribution. The raw data
    from Fig.~\ref{fig:age-mass} (Bottom) has been multiplied with the
    inverse of the contribution matrix (Eq.~\ref{eq:contribution})}

    \label{fig:agemasscor}
\end{figure}

The corrected observations based on Eq.~\ref{eq:correct} are shown in
Fig.~\ref{fig:agemasscor}. The burst at between 50-70 Myr is less
pronounced, but still present. The differences with the uncorrected
observations (Fig.~\ref{fig:age-mass}, Bottom) are small, and we
therefore conclude that our age-fitting method (3DEF) and our applied
selection affect is not severely affecting our age-mass diagrams in a
systematic way. Especially, there is no large systematic shift from
old to young clusters or the other way around. We therefore conclude
that we can use the uncorrected data as well as the corrected data to
compare with the synthetic cluster populations in
\S~\ref{sec:models}. In \S~\ref{subsec:fit-disruption} we will show
that both the corrected as the uncorrected data give the same results
when fitting the analytical models to the data.

\section{Exploration of the parameters which determine the disruption time}
\label{sec:parameters}

\subsection{Cluster initial mass}
\label{subsec:initialmass}

If cluster relaxation drives the evaporation of clusters, then the
more massive clusters live longer than their low mass
counterparts. Boutloukos \& Lamers (\cite{boutloukos03}) have proposed
an empirical way to determine the dependence of the cluster disruption
time on the initial cluster mass, assuming a power-law dependence
of the disruption on the cluster mass

\begin{equation}
t_{\rm dis} = t_4\,(M_i/{10^4\,\msun})^{\gamma}
\label{eq:famous-tdis-eq}
\end{equation}
where $t_4$ is the disruption time of a $10^4 \msun$ cluster, $M_i$ is
the initial mass of the cluster and $\gamma$ is a dimensionless
index. The value of $\gamma$ was determined by measuring the slope of
the age and mass distributions of a cluster population. Their mean
value for $\gamma$ based on four different galaxies was $<\gamma> =
0.62 \pm 0.06$. In a recent study by Lamers, Gieles \& Portegies Zwart
(\cite{lamers05a}) these observational results are compared with
results of $N$-body simulations. The value of $\gamma$ can be
explained by tidally driven relaxation and was confirmed to be
0.62 by $N$-body simulations. The explanation for this is that the
disruption time of a cluster in a tidal field, depends on the
relaxation time ($t_{\rm rel}$) and the crossing time ($t_{\rm cr}$)
of the cluster as (Baumgardt \cite{baumgardt01}):

\begin{equation}
t_{\rm dis} \propto t_{\rm rel}^{x}\times t_{\rm cr}^{1-x}
\label{eq:trel_tcr}
\end{equation}
Using the expression for $t_{\rm rel}$ and $t_{\rm cr}$ from Spitzer
(\cite{spitzer87}), this implies: $t_{\rm dis} \propto \beta(N/\ln
N)^{x}$. Baumgardt \& Makino (\cite{baumgardt03}) found two
combinations of $\beta$ and $x$, depending on the concentration of the
clusters.  Lamers, Gieles \& Portegies Zwart (\cite{lamers05a}) showed
that for both combinations $t_{\rm dis}$ could be well approximated with
$t_{\rm dis} \propto N^{0.62}
\propto M^{0.62}$, where $N$ is the number of stars in the clusters
and $M$ is the total mass of the cluster. King (\cite{king58}) already
had theoretical arguments to expect that the lifetime of clusters
should depend on the mass as $t_{\rm dis} \propto M^{2/3}$. The
agreement between observations and $N$-body simulations was first
noted by Gieles et al. (\cite{gieles04}).  The physical background
of why the disruption time does not scale directly with the relaxation
time is given by Fukushige \& Heggie ({\cite{fukushige00}).

\subsection{Cluster radius}

Young clusters are not only affected by the external tidal field of the
host galaxy, but they also undergo shocks from (giant) molecular
clouds. Both these processes shorten the lifetime of clusters. For
both cases the radius of the cluster is an important parameter in
determining how fast the cluster will disrupt. However, both processes
depend very different on the radius. From Eq.~\ref{eq:trel_tcr} and
the expression for the relaxation time and crossing time it follows
that for the tidally driven relaxation the disruption time depends on
the radius as $t_{\rm dis} \propto r_{\rm h}^{3/2}$. Larger clusters
live longer since they have a longer relaxation time, so it takes more
time for stars to reach the tidal radius and leave the cluster. Spitzer
(\cite{spitzer58}) has shown that the time needed for a cluster to get
unbound due to external shocks, relates to the half mass radius of the
cluster as: $t_{\rm sh} \propto r_{\rm h}^{-3}$, so here larger
clusters live shorter (for isolated clusters).

To see whether the radius of a cluster is an important parameter in
disruption, we use the radii measurements of Paper II. There we
measured the projected half light radius (or effective radius) $r_{\rm
eff}$, which relates to the half mass radius as: $r_{\rm eff} =
3/4\,r_{\rm h}$ (Spitzer \cite{spitzer87}). We make a number density
plot of $r_{\rm eff}$ vs. age for all clusters
(Fig.~\ref{fig:radius-age}). There are clearly no old clusters with
large radius, while the opposite is expected due to the size-of-sample
effect (Hunter et al.~\cite{hunter03}). This suggests that large
clusters are disrupted preferentially. This suggests that shocks may
be the dominating disruption effect. However, when a large fraction of
the clusters is removed, independent of radius, the upper radius would
also go down. This is a result from number statistics: less clusters
in a power law distribution will result in a lower maximum value. So
what really matters here is whether the slope of the radius
distribution changes in time or not. In Paper II it is shown that the
cluster radius distribution of M51 is $N(r)\,dr \propto
r^{-\eta}\,dr$, with $\eta = 2.2 \pm 0.2$. To see how the slope of the
distribution depends on age, we divide our cluster sample in young
(log(Age/yr) $<$ 7.5) and old (log(Age/yr) $>$ 7.5). Dividing the
sample at log(Age/yr) = 7.5 yields two samples of more or less equal
size, which gives similar errros in the fit to both
distributions. When we determine this index $\eta$ for only young
clusters we find $\eta = 2.0 \pm 0.4$ and for old clusters we find
$\eta = 2.5 \pm 0.6$, which is very similar for the value found for
the globular clusters in our Milky Way ($\eta = 2.4 \pm 0.2$, Paper
II). Although the radius distribution seems to get steeper with age,
the errors are too large to place a strong constraint on this. We
therefore do not take the radius into account as a free parameter when
modeling the cluster disruption. Futher studies of M51 with higher
resolution, for example with the {\it Advanced Camera for Surveys
(ACS)}, could shed light on how the radius of clusters affects the
lifetime.

\begin{figure}[t]
    \includegraphics[width=8cm]{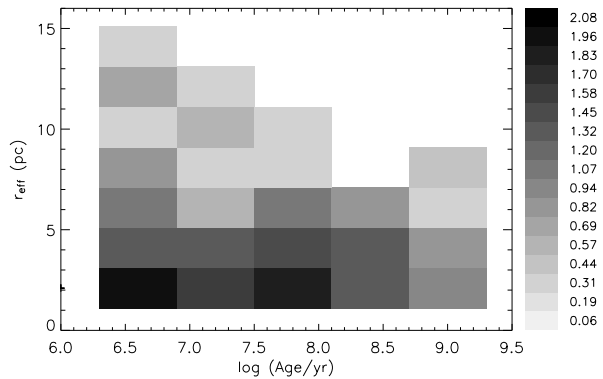} 

    \caption{Number density plot of $r_{\rm eff}$ vs. age. The
    decreasing number going towards larger radii shows the power law
    behavior of the radius distribution. The right hand side shows
    how the grey values correspond to the logarithm of number of clusters.}

    \label{fig:radius-age}
\end{figure}

\subsection{Distance to the galactic center}
Lamers, Gieles \& Portegies Zwart (\cite{lamers05a}) and Baumgardt \&
Makino (\cite{baumgardt03}) have shown that the disruption time is
expected to depend on the galactocentric distance of the cluster, the
orbital velocity in the galaxy and the ambient density of the galaxy as
\begin{eqnarray}
t_{\rm dis} &\propto& R_{\rm G}/V \label{eq:rv}\\
            &\propto&  \rho_{\rm amb}^{-0.5}\label{eq:rho_amb}
\end{eqnarray}
where $R_{\rm G}$ is the distance to the galactic center, $V$ is the
rotational velocity of the cluster in the host galaxy at that distance
and $\rho_{\rm amb}$ is the ambient density of the galaxy at the
location of the cluster. The relation with the ambient density holds
only when a logarithmic potential is assumed. Since we are dealing with
a disk galaxy, it is not so straightforward to derive the ambient
density from the spherically symmetric logarithmic potential. Therefore,
we prefer the relation with the galactocentric distance and the
velocity (Eq.~\ref{eq:rv}). With Eq.~\ref{eq:rv} we are able to
estimate if we would be able to observe a difference in disruption
time at different locations in the galaxy. When we look at clusters
between 1 and 3 kpc and at clusters between 3 and 5 kpc, the average
value of $R_G$ goes up with a factor of 2. The rotational velocity of
M51 increases from 200 $\kms$ to 225 $\kms$ (Rand \cite{rand93}). So
from Eq.~\ref{eq:rv} we expect the disruption time in the two samples
to be different by a factor of 1.8. In Fig.~\ref{fig:nratio} we plot
the ratio of the number of clusters at different ages for the outer
region (3-5 kpc) and inner region (1-3 kpc). Overplotted is the
predicted ratio using Eq.~\ref{eq:mpresent} for disruption timescale
that is different by a factor of 1.8. Just as for the radius
dependence we see that the distance to the galactic center plays a
role, but the data is not sufficient to include it in our analysis.

\begin{figure}[t]
    \includegraphics[width=8cm]{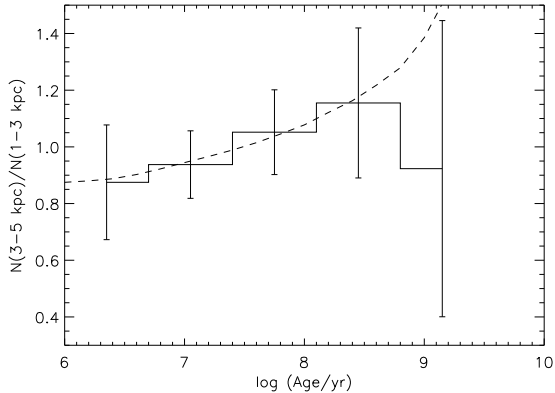}

    \caption{Ratio of the number of clusters ($N$) between 3-5 kpc and
    1-3 kpc as a function of age. Overplotted is a model predicting
    this ratio for two disruption times differing a factor of 1.8,
    based on Eq~\ref{eq:mpresent}.}

    \label{fig:nratio}
\end{figure} 

In conclusion, we see evidence for radius and galactocentric distance
dependent disruption, but the noise is too large to include these
parameters in the models. The mass of the cluster is the most
dominating parameter in the determination of disruption time and in
the remainder of the study we will only use the mass dependence as a
parameter we will vary in the models.

\section{Input parameters for modeling the cluster population of M51}
\label{sec:explanation}

So far, analytical models for finding the cluster disruption time have
assumed that clusters were formed with a constant CFR, as is probably
the case for Galactic open clusters (Boutloukos \& Lamers
\cite{boutloukos03}; Battinelli \& Capuzzo-Dolcetta
\cite{battinelli91}). Lamers~et~al. (\cite{lamers05b}) predicted
the age distribution of open clusters. In the case of M51, we
have age and mass information available for each cluster, so
predictions can be done for age and mass. In addition, assuming a
constant CFR for M51 might be an oversimplification of the situation,
since the galaxy is in interaction with NGC~5195. In the next sections
we explore a broader parameter space.

Since there are strong arguments to believe that the mass dependence of the cluster
disruption ($\gamma = 0.62$) is constant (Lamers, Gieles \& Portegies Zwart
\cite{lamers05a}), we start by varying only the
constant $t_4$, to be able to compare our results with clusters
gradually loosing mass with the {\it instantaneous disruption}
assumption (Eq.~\ref{eq:famous-tdis-eq}) results of Boutloukos \&
Lamers (\cite{boutloukos03}). Next, a two dimensional parameter search
for $\gamma$ and $t_4$ is performed, to verify the assumed value for
$\gamma$ and to study the dependence of $t_4$ on the value of
$\gamma$.

When we have a first estimate of the disruption time, we will study
how this value changes when we assume that the CFR has been increasing
during the last Gyr or contains bursts at the moments of encounter
with NGC~5195. 


\section{Analytical model for generating a cluster population}
\label{sec:models}

\subsection{Setting up a synthetic cluster population}

The synthetic cluster populations will be created in a similar way as
in \S~\ref{sec:artifacts}. This time however we want to include
realistic input physics, like the cluster IMF and different formation
rates, so creating clusters equally spaced in log(Age/yr) and log($M/\msun$)
will not be adequate.  When creating clusters with a realistic CIMF,
the number of clusters needed to fully sample the CIMF up to 10 Gyr
ago is too high. Therefore each cluster was assigned a weight
depending on the initial cluster mass and its age ($w(t,M_i)$),
proportional to the expected number of clusters formed at each age and
mass. The weight is a function of age and mass, scaled such that the
youngest most massive cluster has a weight of 1
\begin{equation}
 w(t, M_i) = (t/t_{\rm min})\times(\alpha -1 ) \times (M_i/M_{\rm max})^{1-\alpha}
\end{equation}
where $w(t, M_i)$ is the weight assigned to a cluster with age $t$ and
 mass $M_i$, $M_{\rm max}$ is the mass of the most massive cluster in
 the simulation, $t_{\rm min}$ is the age of the youngest cluster in
 the simulation and $\alpha$ is the slope of the mass function. When
 $\alpha$ is chosen 2, i.e. $N(M) \sim M^{-2}$, the weight depends on age and mass simply as:
 $w(t,M_i) \propto t/M_i$. When the simulated clusters are binned, the
 weights of the clusters are counted, yielding a realistic log(Age/yr)
 vs. log($M/\msun$) diagram similar to Fig.~\ref{fig:age-mass} (Bottom). The
 advantage of using points spread equally in log(Age/yr) and log($M/\msun$)
 with weights assigned, is that the number of points per bin is
 constant and that it is very easy to create a lot of populations with
 different formation rates, disruption timescales etc.  in a short time.

 In our case the clusters are given a weight such that after binning
the CIMF has a slope of $\alpha = 2.1$ as found for M51 (Paper I) and
the Galactic open clusters (Battinelli et al.~\cite{battinelli94}) and
using different formation and disruption scenarios
(\S~\ref{subsec:fit-disruption}-\S~\ref{subsec:comparison}).

\subsection{Including stellar evolution and cluster disruption}
Baumgardt \& Makino (\cite{baumgardt03}) have shown that stellar
evolution (SEV) is an important contributor to the dissolution of
young clusters, especially for clusters with low concentrations. They
also confirmed that clusters dissolve with a power-law dependence of
their initial mass as $t_{\rm dis} \sim M_i^\gamma$, where $\gamma =
0.62$, in agreement with the empirical determination by Boutloukos \&
Lamers (\cite{boutloukos03}).  In the latter study instantaneous
disruption after the disruption time was assumed as a first
approximation and they found that the typical disruption time ($t_4$,
see Eq.~\ref{eq:famous-tdis-eq}) varies a lot for different
galaxies. In a recent study (Lamers et al. \cite{lamers05b}) it was
shown that there is a simple analytical description of the mass of a
cluster as a function of time. It takes into account the effect of
mass loss due to stellar evolution, based on the mass loss predicted
by the {\it GALEV} SSP models (Anders et al. \cite{anders03}; Schulz
et al. \cite{schulz02}) and cluster mass loss due to the tidal
fields. The mass of the cluster as a function of time can be well
approximated by

\begin{equation}
M_p(t) = ((M_i\mu_{\rm sev})^\gamma - \gamma\frac{t}{t_0})^{1/\gamma}
\label{eq:mpresent}
\end{equation}
where $M_p(t)$ is the present mass of the cluster as a function of its
age, $M_i$ is the initial mass of the cluster, $\mu_{\rm sev} \equiv
M_p(t)/M_i$ is the fraction of remaining mass after mass loss due to stellar
evolution has been subtracted and $t_0$ relates to $t_4$ as $t_4 =
t_0\times10^{4\gamma}$. The mass as a function of time, according to
the analytical formula, agrees perfectly with the predictions
following from $N$-body simulations. Lamers et al.~(\cite{lamers05b}) have also shown
that with this analytical model the age distribution of galactic open
clusters can be explained very well.

\section{Fitting observed age-mass distribution to predictions}
\label{sec:fits}

\subsection{Determining reduced $\chi^2$ values from 2D fits}
\label{subsec:ppl}

Artificial cluster samples with realistic input physics (e.g. a CIMF,
cluster disruption, bursts etc.) can now be generated and compared
with the observed age and mass number density
distribution. 

After calculating the analytically generated cluster population, the model is
binned into number density plots in the same way as the observed data
(see \S~\ref{subsec:agemass}) taking into account the weights. In
order to compare the simulated (2D) age-mass density plots with the
observations, we use the Poisson Probability Law (PPL) introduced by
Dolphin \& Kennicutt (\cite{dolphin02}) for similar purposes
\begin{equation}
{\rm PPL} = 2\sum_{i = 0}^{N}m_i - n_i + n_i\ln\frac{n_i}{m_i}
\label{eq:ppl}
\end{equation}
where $N$ is the number of bins, $m_i$ is the predicted number by the
analytical model in bin $i$ and $n_i$ is the observed number of
clusters in bin $i$. The value of PPL is similar to the $\chi^2$, in
the sense that lower values imply better fits. We will always divide
the PPL value by the number of bins minus the number of degrees of
freedom, which is equivalent to the reduced $\chi^2$, so the
$\redchi$. We will refer to $\redchi$ when we discuss results of fits.

\subsection{Determining the cluster disruption time assuming a constant formation rate of clusters}
\label{subsec:fit-disruption}

To determine the {\it typical} cluster disruption time, $t_4$, defined
in \S~\ref{sec:models}, we generate a cluster sample with a constant
CFR and then calculate the cluster masses as a function of age
according to Eq.~\ref{eq:mpresent} for various values of $t_4$. 
Here we are interested in the disruption time of clusters that have
survived that first 10$^7$ yr in which the natal cloud is being
removed by stellar winds, therefore we exclude the youngest age bin in
the fits. Fig.~\ref{fig:t4-chi2} shows a clear ${\redchi}$ minimum around $t_4 =
\besttfour$ yr, where the upper and lower errors are defined by
${\redchi}_{\rm , accept} \le {\redchi}_{\rm ,min} + 1$, which is
equivalent to the 1 $\sigma$ error. In addition, we have fitted the
same models but then corrected for age-fitting artefacts
(\S~\ref{subsec:correct}). The shape of this $\redchi$ curve is the
same as for the raw data, though the values are higher. This shows
that the uncertainties of our age-fitting method do not alter the
value found for the disruption timescale.

\begin{figure}[t!]
    \includegraphics[width=8cm]{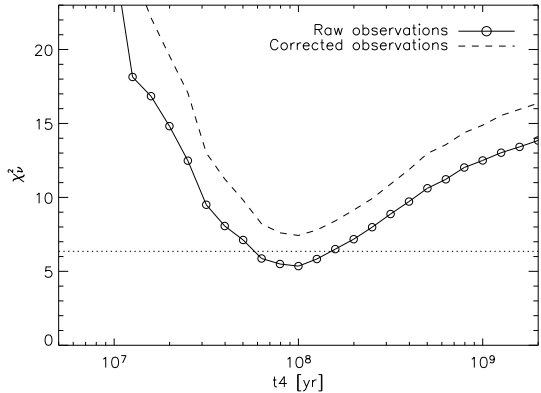} 

    \caption{$\redchi$ values for different disruption times. A clear
    minimum is visible at $t_4 = \besttfour$. The dotted line
    indicates $\redchi_{\rm ,min} + 1$. The dashed line shows a fit to
    the data after correcting for age-fitting artifacts
    (\S~\ref{subsec:correct}). In this simulations we have chosen
    $\gamma$ to be 0.62, based on theoretical arguments
    (\S~\ref{sec:models}).}

    \label{fig:t4-chi2}
\end{figure}

To see how the value of $t_4$ depends on the value of $\gamma$, we
simulate a grid of cluster populations and vary $t_4$ and $\gamma$. A
2D $\redchi$ plot is shown in Fig.~\ref{fig:2d-chi2}. The minimum is
at $\gamma = \bestgamma$ and $t_4 = \besttfourwithgamma$ yr, agreeing
very well with the value of $\gamma = 0.62$, which was stated earlier
based on theoretical arguments and other observational results. The
plot also shows that there is a diagonal bar-shaped minimum for
different combinations of $t_4$ and $\gamma$.  One could argue that
there could be multiple combinations possible, which will yield a
somewhat higher value for $t_4$. The fit however is very sensitive for
the choice of bin size when varying two variables. We excluded the
mass bins higher than $5\times10^5 \msun$, since we probably deal with a
truncation of the mass function. If sampling effects would determine
the upper mass at different ages (Hunter et al. \cite{hunter03}), the
maximum mass should increase much more than we observe in the top
panel of Fig.~\ref{fig:age-mass}. This effect makes the mass function
steeper above log($M/\msun$) $\simeq 5.3$ and therefor that region in
the age/mass diagram is not suitable to fit the (sensitive) mass
dependent disruption. An alternative way to measure $\gamma$ would be to
measure the slopes of the age and mass distribution separately, as was
done in Boutloukos \& Lamers (\cite{boutloukos03}). We fitted these
slopes and found the same value for $\gamma$ as for the 2D fit shown
in Fig.~\ref{fig:2d-chi2}. Again, for the mass, we do not include the
high mass end for similar reasons as mentioned before. This method is
less sensitive for the choice of bin size, since we can fit the slope
of the age and mass distribution independent of the value of the
disruption time. We choose to include the result of the simultaneous
fit of $t_4$ and $\gamma$, because it illustrated nicely how these two
variables relate.

\begin{figure}[t!]
    \includegraphics[width=8cm]{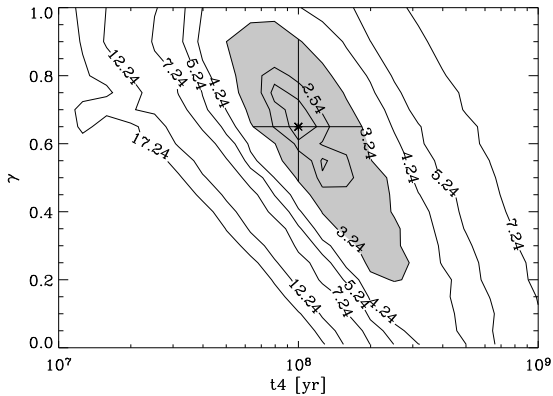} 

    \caption{Two dimensional $\redchi$ plot of $\gamma$ vs. $t_4$. The cross
    and the shaded area indicate the region where $\redchi_{\rm ,min} < \redchi
    < \redchi_{\rm ,min} + 1$ in $t_4$ and $\gamma$.}

    \label{fig:2d-chi2}
\end{figure}

\subsection{The effect of an increasing cluster formation rate}
\label{subsec:var-cfr}

Since NGC~5195 is probably bound to M51 and therefore slowly falling
in (Salo \& Laurikainen \cite{salo00}), one could argue that the short
disruption timescale found in \S~\ref{subsec:fit-disruption} is
actually caused by an increasing cluster formation rate
(CFR). Bergvall et al. (\cite{bergvall03}) have shown that interacting
galaxies such as M51 (i.e. non-merging), can have an increased star
formation rate of the order of a factor of 2-3. We therefore model
different cluster populations with increasing CFR($t$) rates of
various strengths, where we assume that an increasing star formation
rate results in an equally large increase in the CFR($t$). We study
two different models with increasing CFR: 1.) a linear increasing CFR
starting 1 Gyr ago (\S~\ref{subsubsec:linear}); 2.) a CFR that
increases with bursts at the moments of encounter with NGC~5195
(\S~\ref{subsubsec:steps}). Fig.~\ref{fig:cfr-expl} gives a schematic
illustration of how the CFR varies with time for the two models.

\subsubsection{Linearly increasing cluster formation rate}
\label{subsubsec:linear}
In the linear model the CFR($t$) starts to increase 1 Gyr ago, which
is before the moment of the {\it early} close encounter with NGC~5195
(400-500 Myr ago, Salo \& Laurikainen \cite{salo00}). We expect the
CFR to start increasing before the moment of the closest encounter,
since the two galaxies are already interacting before the first
perigalactic passage. We calculate models with different CFR increases
and disruption times. We plot the $\redchi$ values for various values
of CFR($t$ = 0)/CFR($t$ = 10$^9$ yr) and $t_4$ in top panel of
Fig.~\ref{fig:cfr-t4-chi2}. 

\begin{figure}[!t]
    \includegraphics[width=8cm]{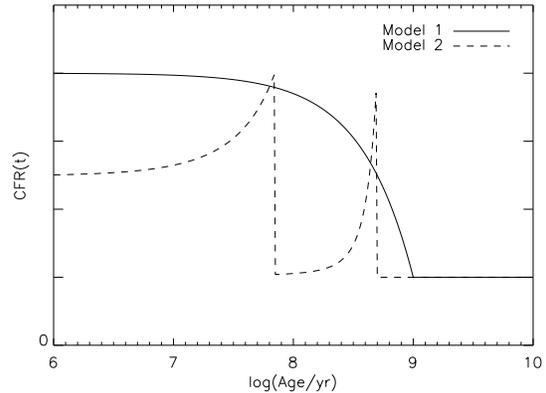}

    \caption{Illustration of the applied CFR increase in
    \S~\ref{subsec:var-cfr} for two different models. Model 1.: The
    CFR is taken to be constant before 1 Gyr ago and then increases
    linear in time until $t=0$. Model 2.: The CFR increases with a
    step at the two moments of encounter with NGC~5195. Here the height
    of the step is the variable.}

    \label{fig:cfr-expl}
\end{figure}
The minimum $\redchi$ value is at $t_4 = \besttfourwithinc $ yr and an
increase in the CFR of $\bestcfrinc$.  For small values of $t_4$ the
equal $\redchi$ lines are vertical. This can be explained by the fact
that {\it if} the disruption time is short, no fingerprints of the
ancient formation rate are present in the current population. They are
simply erased by disruption. the reason that the equal $\redchi$
contours are circular around the minimum, is that the disruption of
clusters depends on the mass of the clusters
(Eq.~\ref{eq:famous-tdis-eq}), unlike an increase of the formation rate.

\subsubsection{Cluster formation rate with bursts}
\label{subsubsec:steps}

An alternative formation scenario would be that the CFR increases with
a burst at the moments of encounter with NGC~5195 and then an
exponential decay in the CFR (see model 2 in
Fig.~\ref{fig:cfr-expl}). We choose the moments of increase at $t =
7\times10^7$ yr and $t = 5\times10^8$ yr ago, based on the results of
Salo \& Laurikainen (\cite{salo00}) and the typical decay time of the
burst is $10^8$ yr (Paper II). The CFR step and $t_4$ are varied in different
models. 
The bottom panel of Fig.~\ref{fig:cfr-t4-chi2} shows that the lowest $\redchi$ value is at $t_4 =
\besttfourwithstep$ yr. This is a factor of 2 higher than when the
increasing CFR is not taken into account, but it is the same value as
was found for the linear increase in the CFR. The value is still a
factor of 5 lower then predicted by $N$-body simulations (Baumgardt
\& Makino
\cite{baumgardt03}; Lamers, Gieles \& Portegies Zwart
\cite{lamers05a}). The best value for the increase in CFR at the moment of encounter is $\bestcfrstep$. The
latter value agrees very well with what is generally observed for the
increase in star formation rate of interacting galaxies (Bergvall et
al. \cite{bergvall03}). Since one of the bursts is clearly observed
and a linearly increasing CFR is not so physical, we prefer Model~2
above Model~1. In the next section we will compare several properties
of this model with the observations.

\begin{figure}[!t]
    \includegraphics[width=8cm]{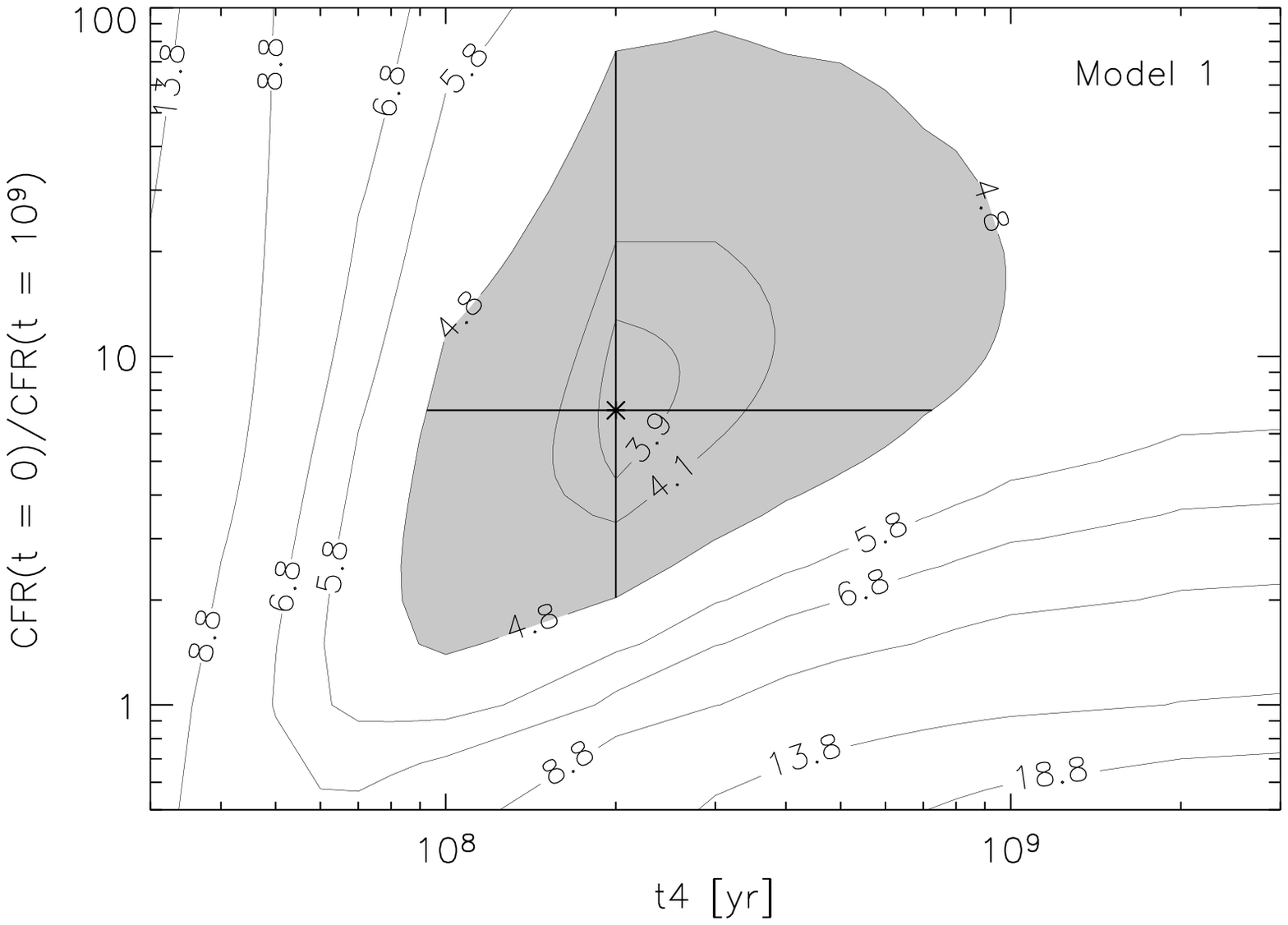} 
    \includegraphics[width=8cm]{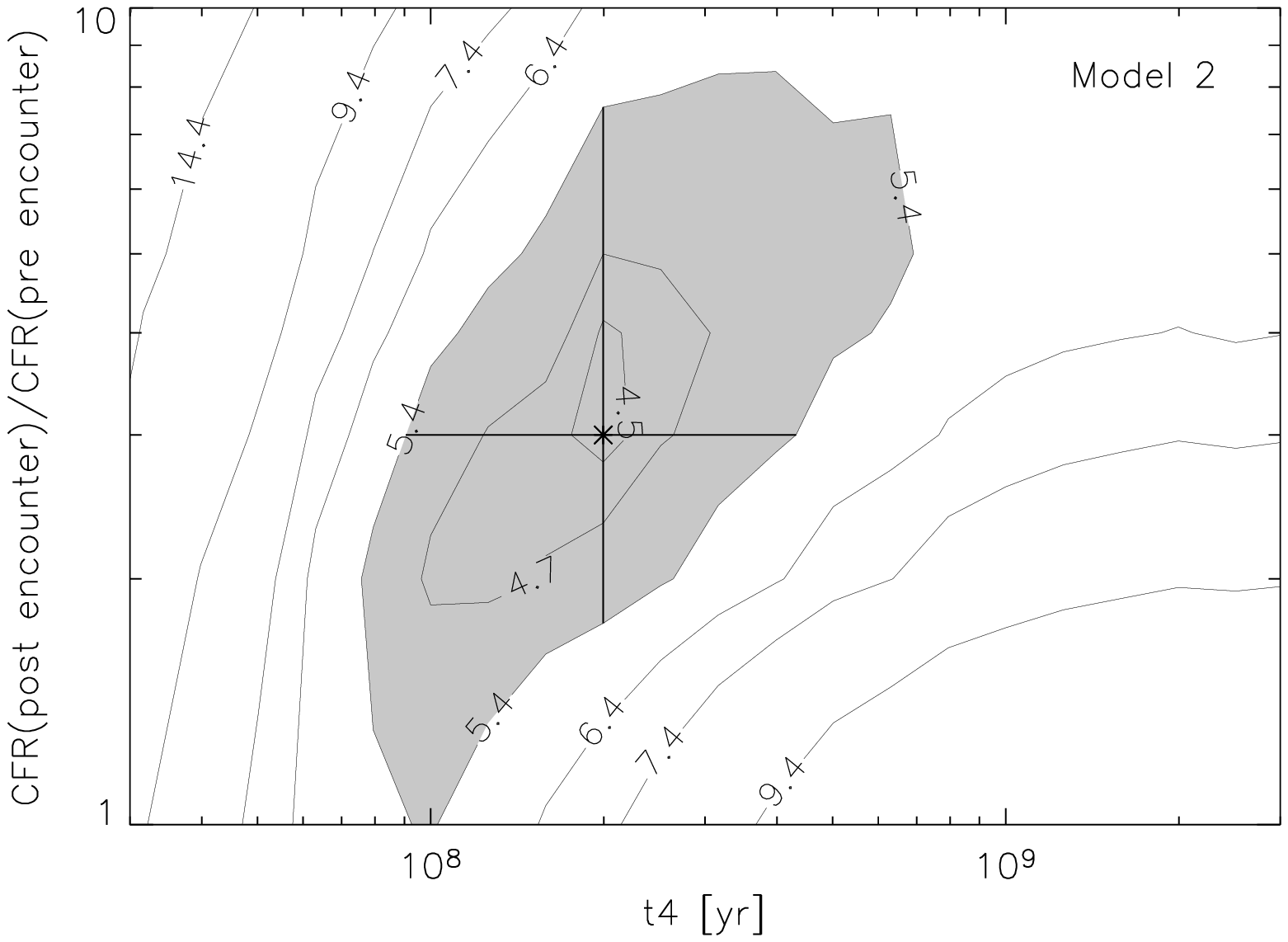} 

    \caption{$\redchi$ values for different combinations of CFR increase and
    disruption time ($t_4$). The top panel shows the results of Model
    1 where the CFR increases linearly between $t$ = 1 Gyr and $t$ =
    present time. The bottom panel shows the result for Model~2 where the CFR
    increases with steps at the moment of encounter with NGC~5195. The
    cross and the shaded area indicate the region where $\redchi_{\rm ,min}
    < \redchi < \redchi_{\rm ,min} + 1$}

    \label{fig:cfr-t4-chi2}
\end{figure}

\subsection{Comparision between the best fit model and the observations}
\label{subsec:comparison}
%


We show a direct comparison between the age-mass diagrams of the best
fit model (\S~\ref{subsubsec:steps}) and the observations in
Fig.~\ref{fig:agemassbestfit}. The densities are scaled such that the
total number of simulated clusters equals the total number of observed
clusters (1152). A few bins in the observations are empty and not
empty in the simulations. The reason for this is that the simulated
cluster sample containts bins with values smaller than 1. Apart from
this, the general trend of grey values in this 2D plot is very similar
in both cases.

\begin{figure}[!t]
    \includegraphics[width=8cm]{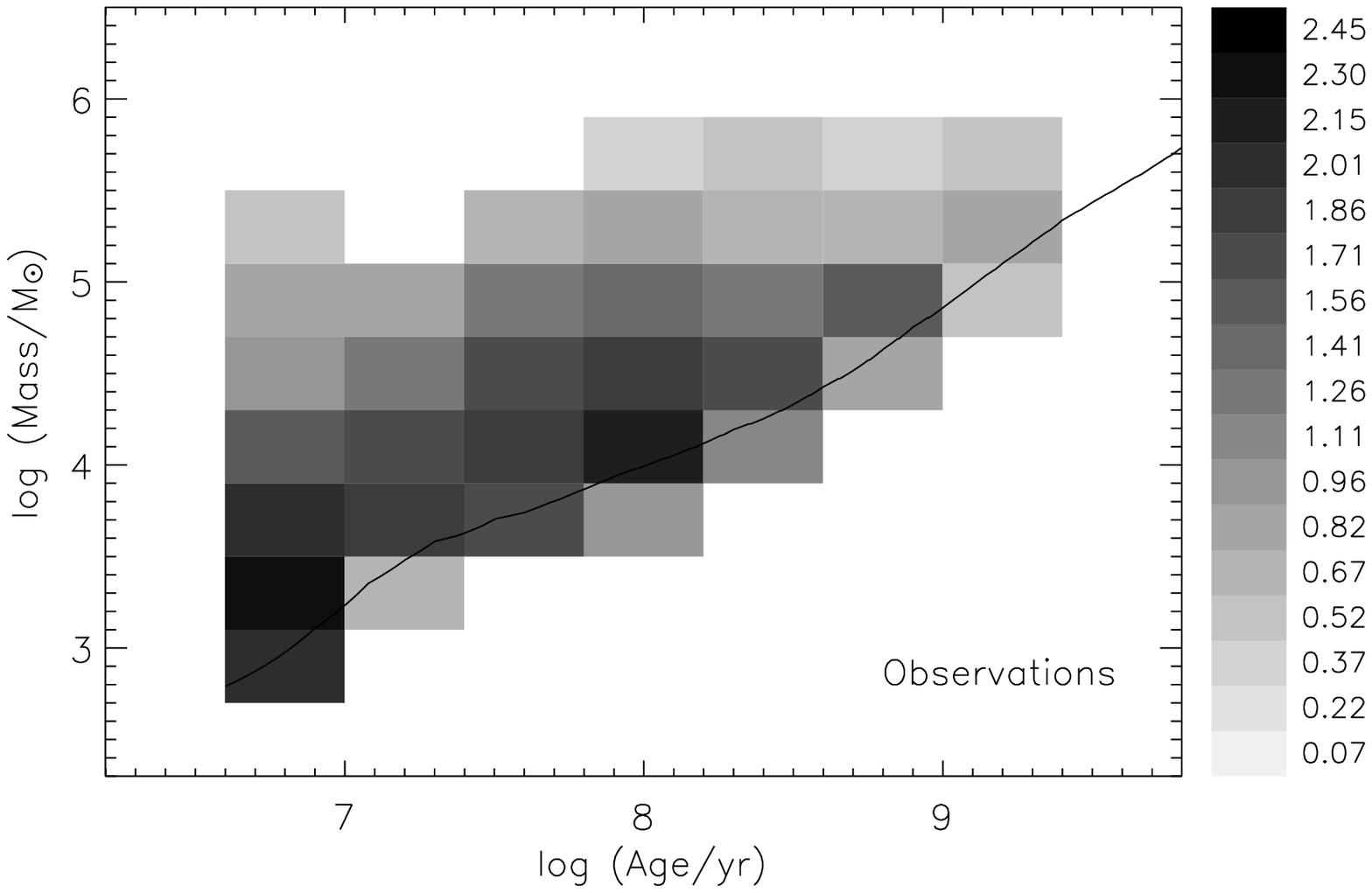}
    \includegraphics[width=8cm]{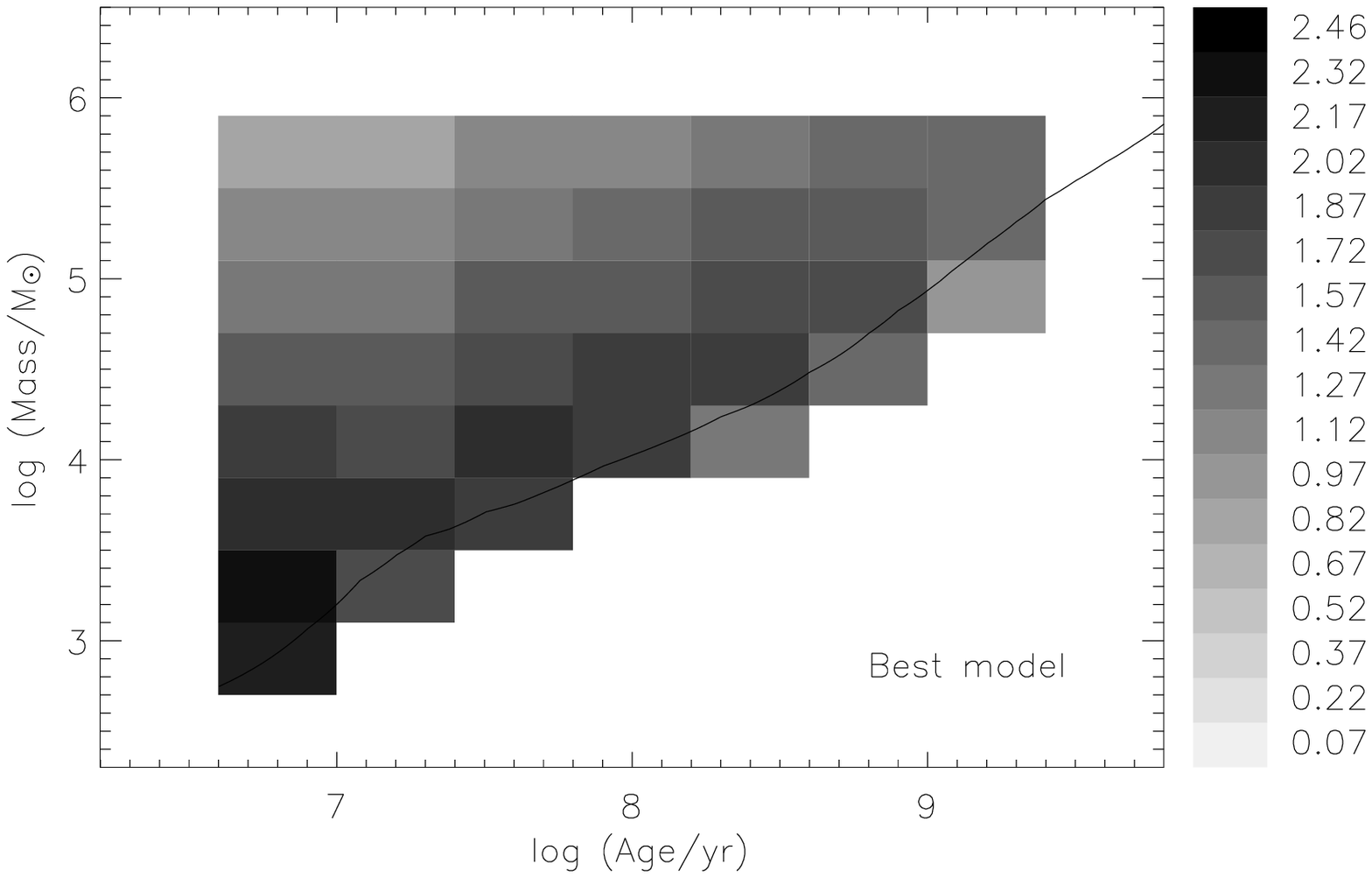}

    \caption{Comparison between the observed ({\bf Top}) and the modeled
    ({\bf Bottom}) age vs. mass number density plots. In both plots the 90\%
    completeness limit of the F439W band is indicated with a line. The
    right hand side shows how the different grey values correspond to
    the logarithm of number. The total number in the simulations is
    scaled to the total number of observed clusters above the 90\%
    completeness limit (1152).}

    \label{fig:agemassbestfit}
\end{figure}

Another interesting property of the observations is the formation
rate. In Paper II we showed the number of clusters at different ages
for different mass cut offs. For clusters with masses higher than
$10^{4.7} \msun$ we get a realistic impression of the cluster
formation rate. This is because we are complete until 1 Gyr for these
masses (see top panel of Fig.~\ref{fig:age-mass}) and because the most
massive clusters are not affected by disruption that much. In
Fig.~\ref{fig:formation} (Top) we show the number of clusters in
different age bins for the observations and the best fit model. The
general trend of the observations is followed very well by the
model. A better way to show the formation rate is to divide each age
bin by the width of the bin. Then we get the number of clusters formed
per unit of time (Myr). This is shown in the bottom panel of
Fig.~\ref{fig:formation}. In this figure the over-density of young
clusters (log(Age/yr) $<$ 7) is more obvious and the burst at
$7\times10^7$ year is better visible. The first burst of cluster
formation ($5\times10^8$ years ago) is not visible anymore, since
clusters with these ages are already affected by the (short)
disruption time. This reinforces that it is very hard to detect
variations in the cluster formation rate when the disruption time is
that short. The largest difference is seen for the bin with
log(Age/yr) = 7 and 8.25. The model predicts in these bins more
clusters than are observed.  This can be explained by fitting
artefacts which yield an (unphysical) underdensity of clusters
(\S~\ref{sec:artifacts}). The model is still within the 3 $\sigma$
error of the observations however.

\begin{figure}[h]
    \includegraphics[width=8cm]{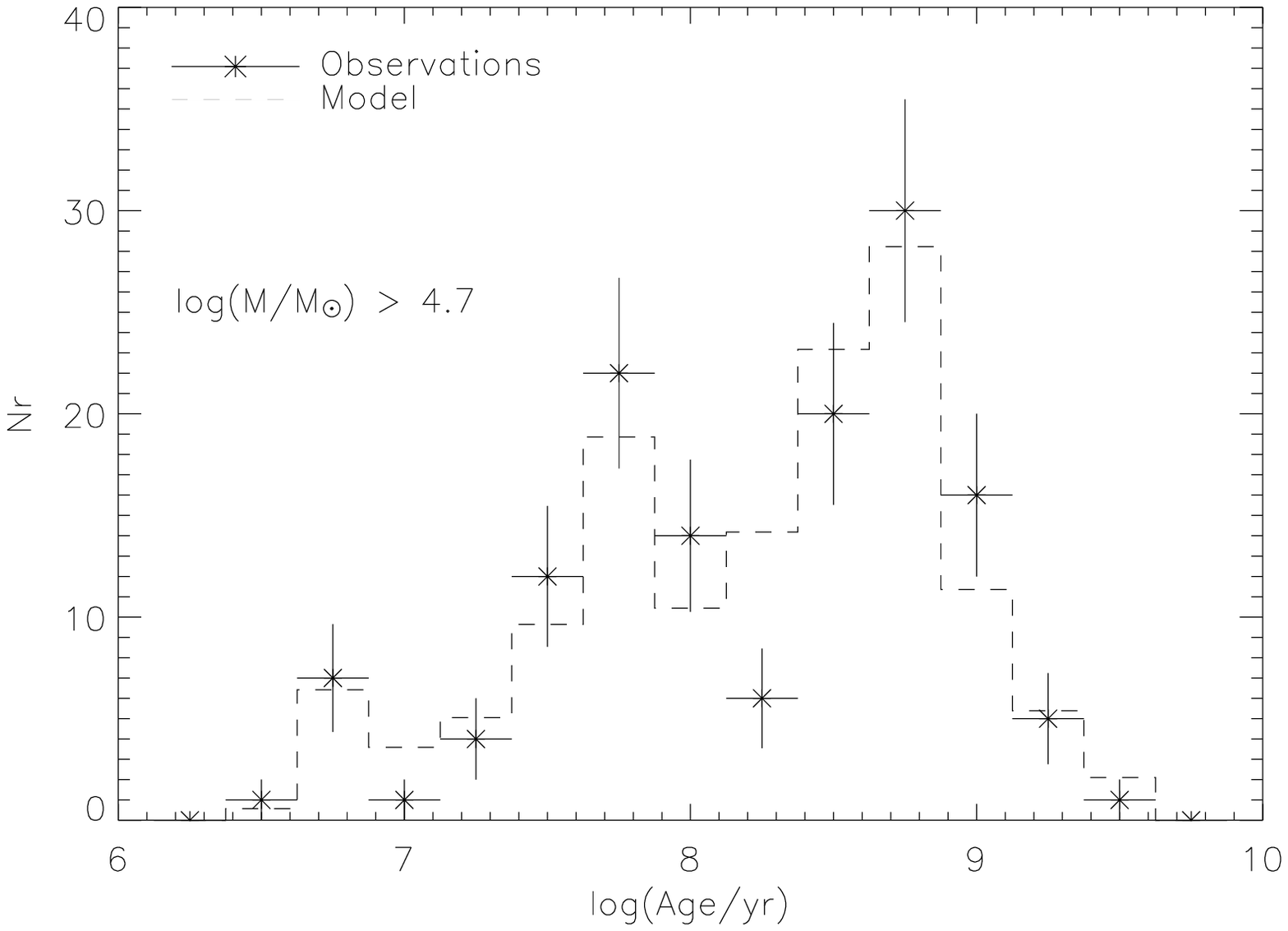} 
    \includegraphics[width=8cm]{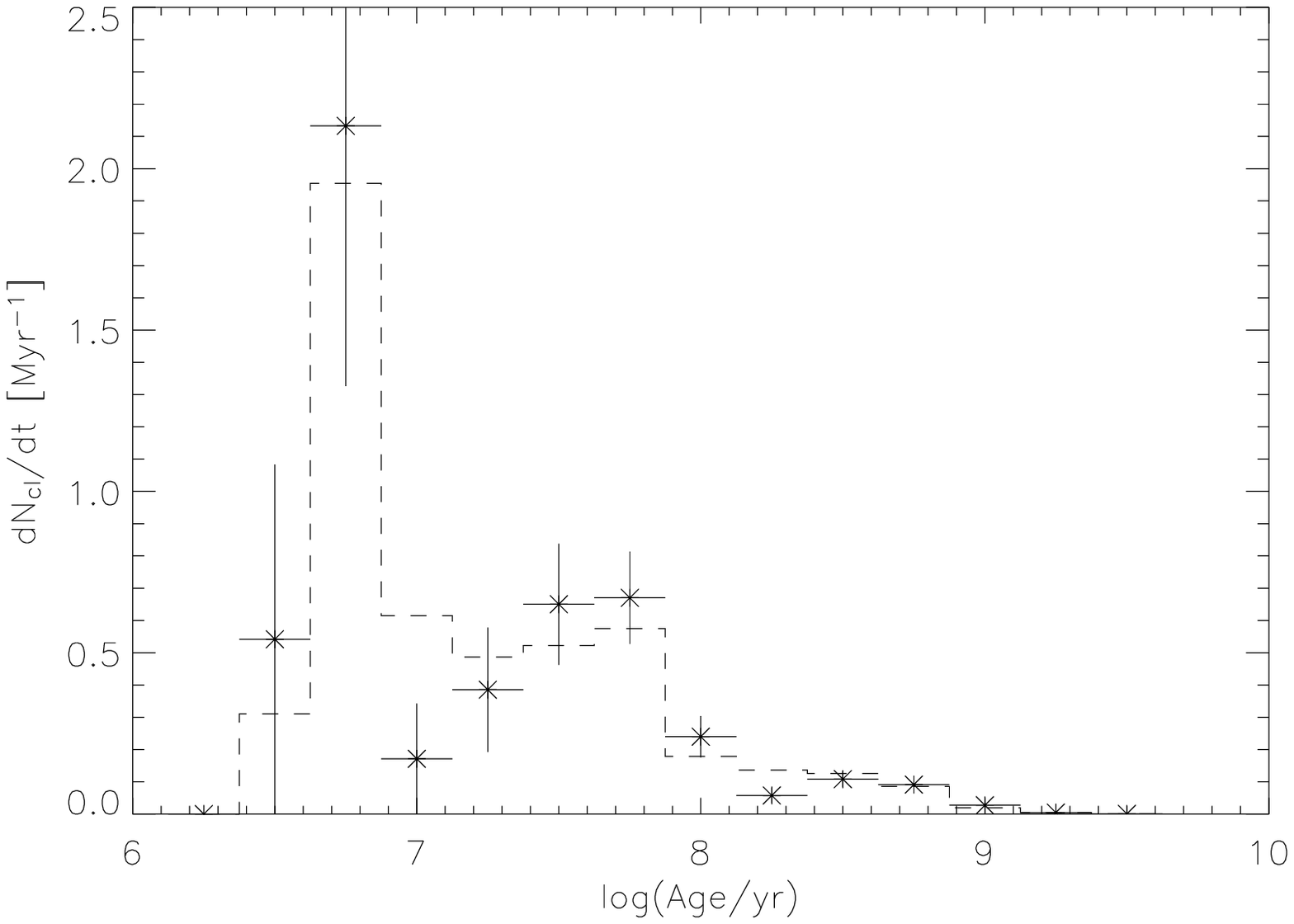} 

    \caption{Comparison of the age distribution of clusters with
    masses larger than $10^{4.7} \msun$ and the one derived from the
    best fit model. {\bf Top:} Number of clusters per age bin. {\bf
    Bottom:} The cluster formation rate (number per Myr).}

    \label{fig:formation}
\end{figure}

%
%
\section{Implication of the derived disruption time}
\label{sec:implications}
We have shown that the clusters disruption time for a typical cluster
with mass of $10^4 \msun$ is around $10^8$ years in M51. When
increasing formation rates are taken into account the disruption time
increases with a factor of 2. This is significantly longer than
Boutloukos \& Lamers (\cite{boutloukos03}) found for clusters in a
smaller region of M51 ($t_4 = 4\times10^7$ yr). This could be because
they did not seperate the dissolution due to infant mortality rate
from the evaporation by the tidal field from the galaxy. Clusters with
ages younger than $10^7$ year are not taken into account in this
study, since they are affected by the dissolution due to the removal
of primordial gas. Theoretical predictions show that clusters of $10^4
\msun$ in a tidal field of the strength of M51 should have an
disruption time of about $10^9$ years (Lamers, Gieles \&
Portegies Zwart \cite{lamers05a}; Baumgardt
\& Makino \cite{baumgardt03}). This value is found from observations of  clusters in
the solar neighborhood (Boutloukus \& Lamers
\cite{boutloukos03}). What causes the clusters in M51 to dissolve
about 5 times faster than predicted? A few
effects that have not been incorporated in the $N$-body models that
predict the disruption times in tidal fields are:

\begin{enumerate}
\item {\bf Variations in the stellar IMF}. When clusters are formed with a
so-called top heavy IMF as is observed in the starburst galaxy M82
(Smith \& Gallagher \cite{smith01}), clusters will disperse much
faster, since the disruption time depends on the number of stars in
the clusters as $t_{\rm dis} \propto N^{0.62}$
(\S~\ref{subsec:initialmass}). Suppose the stellar IMF starts at 1
$\msun$ in stead of $0.1 \msun$, then the number of stars for a given
cluster mass will be about a factor 10 lower. This will make the
disruption time a factor $10^{0.62} \simeq 4$ lower. This would nicely
explain the factor 5 difference in disruption time we observe.

\item {\bf External perturbations}. The $N$-body models of Baumgardt \& Makino
 (\cite{baumgardt03}) calculated the disruption time of clusters in a
 smooth external potential from the host galaxy. In reality, the cluster
 will also experience additional external perturbations, for example the
 encounters with molecular clouds. The clusters in our sample are in
 the inner 5 kpc of the galaxy, where most of the giant molecular
 clouds reside (Henry et al. \cite{henry03}; Kuno~et~al.~\cite{kuno95}). The encounters with
 molecular clouds can speed up the disruption of clouds significantly
 (e.g. Terlevich \cite{terlevich87}; Theuns \cite{theuns91}).

\item {\bf Out of equilibrium formation of clusters}. All clusters in the $N$-body
 models start in virial equilibrium and in tidal equilibrium with the
 host galaxy. Kroupa (\cite{kroupa04}) has shown that after the gas
 removal phase, the clusters are not in virial equilibrium anymore
 and the outer parts of the cluster have expanded. It will be easier
 to dissolve these clusters than when all stars are in tidal
 equilibrium and within the tidal radius imposed by the host galaxy.

\item {\bf Variations in the central concentration}. The $N$-body models of
Baumgardt \& Makino (\cite{baumgardt03}) start clusters with
concentration values of $W_0$ = 5-7. This is the average concentration
of globular clusters in our Milky Way (Harris
\cite{harris96}). When clusters start with much smaller concentration,
the core of the clusters is less compact and the cluster will be more
vulnerable for external perturbations. The concentration of clusters
in M51 can not be determined due to lack of resolution, but we know from
young open clusters in the Milky Way that they have much smaller
concentration indices than the globular clusters (Binney \& Tremaine
\cite{binney87}).

\end{enumerate} 
So far the clusters in M51 have not been checked for variations in the
IMF in the way that it has been done for clusters in other galaxies
(e.g. Smith \& Gallagher \cite{smith01};
Larsen~et~al.~\cite{larsen04};
Maraston~et~al.~\cite{maraston04}). Also, no $N$-body experiments have
been performed including the effect of a tidal field and perturbations
by giant molecular clouds. Argument 3 and 4 are based upon unknown
observables of young clusters and they could hold for clusters in
other galaxies as well. 


When the disruption of clusters is indeed as short as we derived,
young massive clusters ($M_i \simeq 10^6 \msun$) will not survive
longer than 3.5$\times10^9$ yr. This means that the disk of M51 is not
the right location for young globular clusters to survive over a Hubble
time.

\section{Conclusions}
\label{sec:conclusions}

We have compared the cluster population of M51 with theoretical
predictions including evolutionary mass loss, cluster disruption,
variable cluster formation rate and a magnitude limit. The age
vs. mass diagrams of the observed cluster populations are binned to
acquire two dimensional number density plots, which can be compared
with simulated cluster samples. The results can be summarized as
follows:

\begin{enumerate}

\item Artifacts introduced by our age-fitting routine do not systematically
 bias our sample towards young or old clusters. We present a method to
 correct observations of a cluster population for artifacts introduced
 by the age-fitting method applied.

\item The size of the largest cluster decreases with age, from 15 pc for
clusters with log(Age/yr) $<$ 7 to 10 pc for clusters with ages around 1
Gyr. In addition, the slope of the radius distribution seems to gets
steeper in time: $\eta = 2.0 \pm 0.4$ for clusters younger than
log(Age/yr) = 7.5 and $\eta = 2.5 \pm 0.6$ for clusters with log(Age/yr)
$>$ 7.5. Both these results seem to suggest that smaller clusters have
a larger chance to survive. However, the radius distribution of
globular clusters in our Milky Way is very similar to these values
found: $\eta = 2.4 \pm 0.2$. Samples with higher spatial resolution
and more clusters are needed to study the radius dependence.

\item There are more old clusters at larger distances from the galactic
center. The ratio of the number of clusters in the outer parts of the
galaxy (3-5 kpc) over the number of clusters in the inner part (1-3
kpc) per age bin increases with a factor of 1.8 in age (from log(Age/yr)
= 6.5 to log(Age/yr) = 8.5), which is to be expected since the disruption
time depends on the distance to the galactic center.

\item  Assuming that the cluster disruption time depends on the initial mass of
the cluster as $t_{\rm dis} \propto M_i^{\gamma}$, and using $\gamma =
0.62$ based on theoretical and observational studies (Lamers, Gieles
\& Portegies Zwart \cite{lamers05a}), we find a typical disruption for
a 10$^4 \msun$ cluster of $t_4 = \besttfour$ yr, where we assumed a
constant cluster formation rate.

\item When $\gamma$ and $t_4$ are varied together, the value found for
 $\gamma$ is similar to that predicted by Lamers, Gieles \& Portegies
 Zwart (\cite{lamers05a}) based on observational and $N$-body
 studies. A value of $\gamma = \bestgamma$ and $t_4 =
 \besttfourwithgamma$ yr are the best combination.

\item We studied the degeneracy between formation increase and disruption.
Models where the cluster formation rate increases linearly in time do
not affect the disruption timescale much ($t_4$ gets a factor 2
higher). When we include bursts at the moments of encounter with
NGC~5195, the typical disruption time is also a factor of 2 higher.

\item When clusters of 10$^4 \msun$ are disrupted within $2\times10^8$
years, and considering the power-law dependence of the disruption time
scale with the initial cluster mass, even clusters with a mass of
$10^6 \msun$ will not survive longer than 3.5 Gyr. This means that the
disk of M51 is not a preferred location to form a new generation of
globular clusters. This might explain why there so far are no old ($>$
Gyr) massive ($> 10^6 \msun$) clusters known in the disks of spiral
galaxies, although they are still forming (e.g. Westerlund 1 in the
Galactic disk (Clark \& Negueruela \cite{clark04}) and the young
globular cluster in NGC~6946 (Larsen et al. \cite{larsen01})).

\end{enumerate}

        

\bibliographystyle{alpha}
\bibliography{../bib/astroph.bib,../bib/phd.bib,../bib/mark.bib}

\begin{thebibliography}{}

\bibitem[2003]{anders03} Anders, P.,~\& Fritze-v.~Alvensleben, U.\ 2003, \aap, 401, 1063

\bibitem[2004]{anders04} Anders, P., Bissantz, N., Fritze-v.~Alvensleben,
U., \& de Grijs, R. 2004, \mnras, 347, 196
\bibitem[2003]{bastian03} Bastian, N., \& Lamers, H. J. G. L. M. 2003, in Extragalactic Globular Cluster Systems, ed. M. Kissler-Patig, ESO Astrophysics Symposia (Springer: Berlin), 28
\bibitem[2005]{bastian05} Bastian, N., Gieles, M., Lamers, H.~J.~G.~L.~M., Scheepmaker,
 R.A., \& de Grijs, R.  2005, \aap, 431, 905 (Paper II)

\bibitem[1994]{battinelli94} Battinelli, P.,  Brandimarti, A., \&  Capuzzo-Dolcetta, R. 1994, \aaps, 104, 379

\bibitem[1991]{battinelli91}Battinelli, P., \& Capuzzo-Dolcetta, R., 1991, \mnras,  249, 76
\bibitem[2001]{baumgardt01}Baumgardt, H.  2001, \mnras, 325, 1323

\bibitem[2003]{baumgardt03} Baumgardt, H.,~\& Makino, J.\ 2003, \mnras, 340, 227 
\bibitem[1987]{binney87} Binney J., \& Tremaine S. 1987, Princeton, NJ, Princeton University Press, 747

	
\bibitem[2003]{bergvall03} Bergvall, N., Laurikainen, E., \& Aalto, S. 2003 \aap, 405, 31

\bibitem[2003]{bik03} Bik, A., Lamers, H.~J.~G.~L.~M., Bastian, N.,
Panagia, N., \& Romaniello, M. 2003, \aap, 397, 473 (Paper I)
%
\bibitem[2003]{boutloukos03} Boutloukos, S.G., \& Lamers, H.~J.~G.~L.~M. 2003, \mnras, 338, 717 

\bibitem[1998]{brodie98}{Brodie}, J.~P., {Schroder}, L.~L., {Huchra}, J.~P., 
	{Phillips}, A.~C., {Kissler-Patig}, M., \& {Forbes}, D.~A. 1998, \aj, 116, 691
%


\bibitem[2004]{clark04}Clark, J. S., \& Negueruela, I. 2004 \aap,  413, 15
\bibitem[2005]{degrijs05} de Grijs, R., Anders P., Lamers H.~J.~G.~L.~M., Bastian N.,  Fritze-v. Alvensleben, U., 
Parmentier G., Sharina M.E., \& Yi S. 2004, \mnras, accepted for publication

\bibitem[2002]{dolphin02} Dolphin, A.~E.,~\& Kennicutt, R.~C.\ 2002, \aj, 124, 158 


%
%
%
%
\bibitem[2000]{fukushige00} Fukushige, T., \& Heggie, D. C. 2000, \mnras, 318, 753
\bibitem[2001]{geyer01} Geyer, M.P., \& Burkert, A. 2001, \mnras, 323, 988-994
\bibitem[2004]{gieles04} Gieles, M., Baumgardt, H., Bastian, N., \& Lamers, H.~J.~G.~L.~M. 2004, ASP Conf. Ser. vol. 322, p. 481, {\it ''The formation and evolution of massive young star clusters''}, eds. H.L.G.L.M. Lamers, L.J. Smith and A. Nota, ASP San Francisco.
\bibitem[2005]{gieles05} Gieles, M., Larsen, S.~S., Bastian, N., \& Stein, I.~T. 2005, \aap, submitted
\bibitem[1996]{harris96} Harris, W.E. 1996, \aj, 112, 1487

\bibitem[2003]{henry03}Henry, A.~L., Quillen, A.~C., \& Gutermuth, R. 2003, \aj, 126, 283
\bibitem[1987]{hodge87}Hodge, P. 1987, \pasp, 99, 724
\bibitem[2003]{hunter03} Hunter, D.~A., {Elmegreen}, B.~G., {Dupuy}, T.~J., \&
	{Mortonson}, M. 2003, \aj, 126, {1836-1848}
%
%
%
%
\bibitem[1958]{king58} King, I. 1958, \aj, 63, 306
\bibitem[2004]{kroupa04} Kroupa, P. 2004, \ap, 0412069


\bibitem[1995]{kuno95}Kuno, N., Nakai, N., Handa, T., \& Sofue, Y. 1995, \pasj, 47, 745
\bibitem[2003]{lada03} Lada, C.~J. \& Lada, E.~A. 2003, \araa, 41, 57
\bibitem[2005]{lamers05a}  Lamers, H.~J.~G.~L.~M., Gieles, M., \& Portegies Zwart, S.~F. 2005, \aap, 429, 173
\bibitem[2005]{lamers05b}  Lamers, H.~J.~G.~L.~M., Gieles, M., Bastian, N.,  Baumgardt, H., Kharchenko, N.~V., \& Portegies Zwart, S.~F. 2005, \aap, submitted

\bibitem[2001]{larsen01}Larsen, S. S., Brodie, J. P., Efremov, Y. N., Elmegreen, B. G., Hodge, P. W., 
\&  Richtler, T.  2001, \apj, 556 , 801
\bibitem[2004]{larsen04}Larsen, S.~S., Brodie, J.~P., \& Hunter, D.~A. 2004. \aj, 128, 2295

%
%
%
%
%
%

\bibitem[2004]{maraston04}Maraston, C., Bastian, N., Saglia, R.~P., Kissler-Patig, M., 
Schweizer, F., \& Goudfrooij, P. 2004, \aap, 416, 467
\bibitem[1993]{rand93} Rand, R.~J.\ 1993, \apj, 410, 68

\bibitem[2000]{salo00} Salo, H., \& Laurikainen, E. 2000, \mnras, 319, 377
%
%
\bibitem[2002]{schulz02} Schulz, J., Fritze-v. Alvensleben, U., \& Fricke, K.J. 2002, \aap, 392, 1
\bibitem[2001]{smith01}	Smith, L.~J., \& Gallagher, J.~S. 2001, \mnras, 326, 1027

\bibitem[1958]{spitzer58} Spitzer, L.~J.\ 1958, \apj, 127, 17
\bibitem[1987]{spitzer87}{Spitzer}, L., 1987, Princeton, NJ, Princeton University Press, 1987, p. 40 + p. 115
%
\bibitem[1987]{terlevich87} Terlevich E., 1987, \mnras, 224, 193
%
\bibitem[1991]{theuns91} Theuns T., 1991, Memorie della Societa Astronomica Italiana, 62, 909


%
%
%
%
%
\end{thebibliography}

\end{document}